\documentclass[twocolumn,preprintnumbers,amsmath,amssymb,10pt,a4paper,superscriptaddress,nofootinbib,showpacs]{revtex4}
\pdfoutput=1
\usepackage{geometry}
\usepackage{color}

\usepackage{graphicx}
\usepackage{dcolumn}
\usepackage{bm}
\newcommand{\be}{\begin{equation}}
\newcommand{\ee}{\end{equation}}
\newcommand{\bd}{\begin{displaymath}}
\newcommand{\ed}{\end{displaymath}}
\newcommand{\BE}{\begin{eqnarray}}
\newcommand{\EE}{\end{eqnarray}}

\newcommand{\id}{{\rm 1\!\!I}}

\newcommand{\bk}{\ensuremath{\mathbf{k}}}

\newcommand{\br}{\ensuremath{\mathbf{r}}}

\newcommand{\bx}{\ensuremath{\mathbf{x}}}
\newcommand{\by}{\ensuremath{\mathbf{y}}}

\newcommand{\bn}{\ensuremath{\mathbf{n}}}

\newcommand{\hx}{\widehat{x}}
\newcommand{\hy}{\widehat{y}}

\newcommand{\D}{{\cal D}}

\newcommand{\boldxi}{{\mbox{\boldmath $\xi$}}}
\newcommand{\boldeta}{{\mbox{\boldmath $\eta$}}}

\newcommand{\boldpsi}{{\mbox{\boldmath $\psi$}}}
\newcommand{\boldphi}{{\mbox{\boldmath $\varphi$}}}

\newcommand{\avg}[1]{\left\langle{#1}\right\rangle}

\newcommand{\olK}{\overline{K}}
\newcommand{\olQ}{\overline{Q}}

\newcommand{\cB}{{\cal B}}

\begin{document}

\title{Stochastic processes with distributed delays: chemical Langevin equation and linear-noise approximation}

\author {Tobias Brett}
\email{tobias.brett@postgrad.manchester.ac.uk }
\affiliation{Theoretical Physics, School of Physics and Astronomy, The University of Manchester, Manchester M13 9PL, United Kingdom}

\author{Tobias Galla}
\email{tobias.galla@manchester.ac.uk}
\affiliation{Theoretical Physics, School of Physics and Astronomy, The University of Manchester, Manchester M13 9PL, United Kingdom}


 \begin{abstract}
We develop a systematic approach to the linear-noise approximation for stochastic reaction systems with distributed delays. Unlike most existing work our formalism does not rely on a master equation, instead it is based upon a dynamical generating functional describing the probability measure over all possible paths of the dynamics. We derive general expressions for the chemical Langevin equation for a broad class of non-Markovian systems with distributed delay. Exemplars of a model of gene regulation with delayed auto-inhibition and a model of epidemic spread with delayed recovery provide evidence of the applicability of our results.
\end{abstract}
\maketitle

{\em Introduction.} The theory of discrete Markov processes is well established, and has found applications in a variety of disciplines, including biology, chemistry, physics, evolutionary dynamics, finance and the social sciences \cite{markov}. The standard mathematical treatment is the chemical master equation \cite{gardiner}. Exactly soluble problems are an exception, although they include notable examples such as the voter model \cite{FK96}. The majority of Markovian systems can only be analysed using approximative schemes, such as the van Kampen or the Kramers-Moyal expansions \cite{vk, km,gardiner}. Truncating these expansions after sub-leading order leads to a Gaussian approximation, the so-called chemical Langevin equation \cite{cle}. When linearised about a deterministic trajectory this is known as the linear-noise approximation (LNA) in chemistry and biology \cite{vk}. The Gaussian approximation and the LNA provide an important starting point for further analytical studies and for efficient 
simulations \cite{cle,lnasim}. Analytical approaches of this type have been applied to a wide range of problems \cite{various}, and for many model systems they reflect the current state of play. Schemes going beyond Gaussian order are only currently being constructed \cite{grima}.

The purpose of our work is to develop a comprehensive picture of the LNA for interacting-particle systems with delay. The time evolution of delay systems depends on the prior path the system has taken. Existing approaches include Fokker-Planck equations \cite{frank} and time-scale separation \cite{bratsun}. The system-size expansion to first order has been carried out in \cite{bratsun,galla} for a model with one fixed delay time. Recent work \cite{lafuerza} has extended these approaches to systems with distributed delays. These are recognised as more realistic than models with constant delays \cite{distrib,monk,andyalan}, but a comprehensive formalism is still lacking. 

Most existing work on stochastic delay models is based on extensions of the master equation for delay systems. We take a different approach and choose a generating function description of entire paths of the dynamics \cite{msr}. This formalism is originally due to Martin, Siggia, Rose, Janssen and De Dominicis (MSRJD), and it is not to be confused with a generating function approach to solving master equations. The MSRJD-formalism removes the need for a master equation altogether. This provides a new perspective on stochastic delay systems, and, we think, it allows one to carry out the LNA more naturally and systematically. As a consequence we are able to derive an explicit Gaussian approximation for a broad class of delay models, ready to be applied to problems with delay dynamics in a number of fields. 

{\em Generating functional approach to delay systems.}  Consider a reaction system with $S$ types of particles, $\alpha=1,\dots,S$. The state of the system is characterised by $\bn(t)=(n_1(t),\dots,n_S(t))$, where the integer $n_\alpha(t)$ indicates the number of particles of type $\alpha$ at time $t$. The dynamics occurs via $R$ possible reactions, $i=1,\dots,R$. The rate with which reaction $i$ fires is denoted by $T_i(\bn)$.  Each reaction can result in a change of particle numbers at the time the reaction is triggered, and at a later time. The latter aspect reflects the delay interaction. We write $v_{i,\alpha}$ for the change in the number of particles of type $\alpha$ at the time a reaction of type $i$ is triggered. Additionally when a reaction of type $i$ fires at time $t$ a delay time $\tau>0$ is drawn from a distribution $K_i(\cdot)$.  A further change of particle numbers occurs at time $t+\tau$, indicated by the variables $w_{i,\alpha}^\tau$. This description includes Markovian processes, one then 
has $w_{i,\alpha}^\tau=0$. 

The purpose of expansion methods is to construct Gaussian stochastic differential equations (SDEs) approximating the statistics of the reaction dynamics \cite{gardiner,vk}. These procedures rely on a large parameter, $N$, in most cases a scale setting the number of particles in the system. Time is scaled so that reaction rates are of order $N$, $T_i(\bn)=Nr_i(\bx)$, and relative particle numbers $x_\alpha=n_\alpha/N$ are introduced. An expansion in negative powers of $N$ then leads to an effective SDE for $\bx$, valid in the limit of large, but finite $N$. Equivalent effective SDEs can be obtained using a theorem due to Kurtz \cite{kurtz}. These techniques, however, are only applicable for Markovian systems.

The starting point for our generating function approach is a discretised dynamics. Introducing a time step $\Delta$ we assume  that the number of reactions of type $i$ firing at time step $t$ and with a delayed effect precisely $\tau\in \mathbb{N}\Delta$ time steps later is a Poissonian random variable, $k_{i,t}^\tau$, with mean $Nr_i[\bx(t)]K_i(\tau)\Delta^2$ (see Appendices~\ref{appendix:disc_time} and \ref{appendix:delay_generating}) \cite{remark-1}. We will write ${\cal P}(\bk)$ for their joint distribution, suppressing the dependence on $\bx$. The generating function for the discrete-time process is then given by
\BE
Z[\boldpsi]&=&\sum_{\bk}  \int D\bx ~{\cal P}(\bk) \exp\left(i\Delta\sum_{t,\alpha}\psi_{\alpha,t} x_{\alpha,t}\right)\nonumber \\
&&\times \prod_{t,\alpha} \delta\left[x_{\alpha, t+\Delta}-x_{\alpha,t}-\phi_\alpha(\bk)\right].\label{eq:gf}
\EE
We have here introduced the source term $\boldpsi$ whose role is to generate the moments of the $\{x_{\alpha,t}\}$. The (re-scaled) total change of the number of particles of type $\alpha$ at time step $t$ is given by 
\be 
\phi_\alpha(\bk)=N^{-1}\sum_i\left( k_{i,t}v_{i,\alpha}+\sum_{\tau\geq \Delta} k_{i,t-\tau}^\tau w_{i,\alpha}^\tau\right),
\ee
 where $k_{i,t}=\sum_{\tau\geq\Delta}k_{i,t}^\tau$. By writing the $\delta$-functions in Eq. (\ref{eq:gf}) in their exponential representation, performing the average over the $\{k_{i,t}^\tau\}$, keeping only leading and sub-leading terms in an expansion in powers of $N^{-1}$, and subsequently taking the limit $\Delta\to 0$ a continuous-time generating functional is obtained. These steps are described in detail in Appendix~\ref{sec:delay}. The resulting generating-functional is equivalent to the Gaussian dynamics 
\be\label{eq:gaussian}
\dot x_\alpha=F_\alpha(t,\bx)+N^{-1/2}\eta_\alpha, 
\ee
with $\avg{\eta_\alpha(t)\eta_\beta(t')}=B_{\alpha,\beta}(t,t',\bx)$, and where
\be
\hspace{-0.5em}F_{\alpha}(t,\bx)=\sum_i \left[r_i[\bx(t)]v_{i,\alpha}+\!\!\int_0^\infty\!\!\! d\tau K_i(\tau) r_i[\bx(t-\tau)]w_{i,\alpha}^\tau\right].\label{eq:drift}
\ee
We set $K_i(\tau)=0$ for $\tau<0$, and introduce
\BE
B_{\alpha,\beta}(t,t',\bx)&=&\sum_i\bigg\{\delta(t-t')  \bigg[ r_i[\bx(t)]v_{i,\alpha}v_{i,\beta}\nonumber \\
&&\hspace{-8em}+\int_0^\infty\!\!\! d\tau~ r_i[\bx(t-\tau)]K_i(\tau)w_{i,\alpha}^\tau w_{i,\beta}^\tau \bigg]+ \bigg[ r_i[\bx(t)] K_i(t'-t)\nonumber \\
&&\hspace{-6em} \times v_{i,\alpha}w_{i,\beta}^{(t'-t)}+ r_i[\bx(t')] K_i(t-t') v_{i,\beta}w_{i,\alpha}^{(t-t')}\bigg] \bigg\}.\label{eq:corr}
\EE
Eqs. (\ref{eq:gaussian}, \ref{eq:drift}, \ref{eq:corr}) define the chemical Langevin equation for systems with distributed delay. They are the main result of our paper and provide general expressions for the Gaussian approximation of a wide class of delay systems \cite{remark0, remark1}. These equations allow one to disentangle the contributions of the different reactions to the noise, and they can be used for efficient numerical simulations. The gain in computing time can be significant, see Appendix~\ref{appendix:simulate} for further details. The result of Eqs. (\ref{eq:gaussian}, \ref{eq:drift}, \ref{eq:corr}) is slightly stronger than the LNA \cite{gardiner}, which can be obtained from a straightforward linearisation (see Appendix~\ref{appendix:lna}). The resulting linear dynamics is an important intermediate step for further analytical investigations. In the following we will demonstrate the applicability of this approach. We will use our results to compute the spectra of noise-induced quasi-cycles \cite{mckane} in a model of gene regulation 
and in a model of epidemic spread, both with delay interactions.

{\em Application to a model of gene regulation.} Delays in transcription and translation play an important role in gene regulation. They are considered a potential mechanism for oscillatory behaviour in somitogenesis, giving rise to spatially heterogeneous cellular structures \cite{saga.2001, gene.2003}. Models of these processes have traditionally focused on differential equations, see e.g. \cite{gene.2003}. It is only more recently that intrinsic noise has been included \cite{barrio, bratsun}. This is due to the observation that particle numbers in gene regulatory systems can be small, making deterministic approximations inadequate \cite{rao.2002}. For example noise-driven quasi-cycles go undetected in deterministic models \cite{mckane}. Existing theoretical analyses are limited to models with constant delay periods \cite{bratsun, galla}, we note recent advances \cite{lafuerza}. Our result for systems with distributed delay provides a systematic theoretical framework, and we apply it to the simple model of 
gene regulation described in \cite{barrio,gene.2003}. We consider two types of particles, mRNA molecules, denoted by $M$, and protein molecules, $P$. The stochastic dynamics are given by
\BE
M &\overset{\mu_M}{\longrightarrow}& \emptyset,  \nonumber \\
P &\overset{\mu_P}{\longrightarrow}& \emptyset,  \nonumber \\
M &\overset{\alpha_P}{\longrightarrow}& M + P,   \nonumber \\
\emptyset &\overset{g(n_P), K(\tau)}{\Longrightarrow}& M.  
\EE
\begin{figure}[t!!]
\centerline{\includegraphics[width=0.45\textwidth]{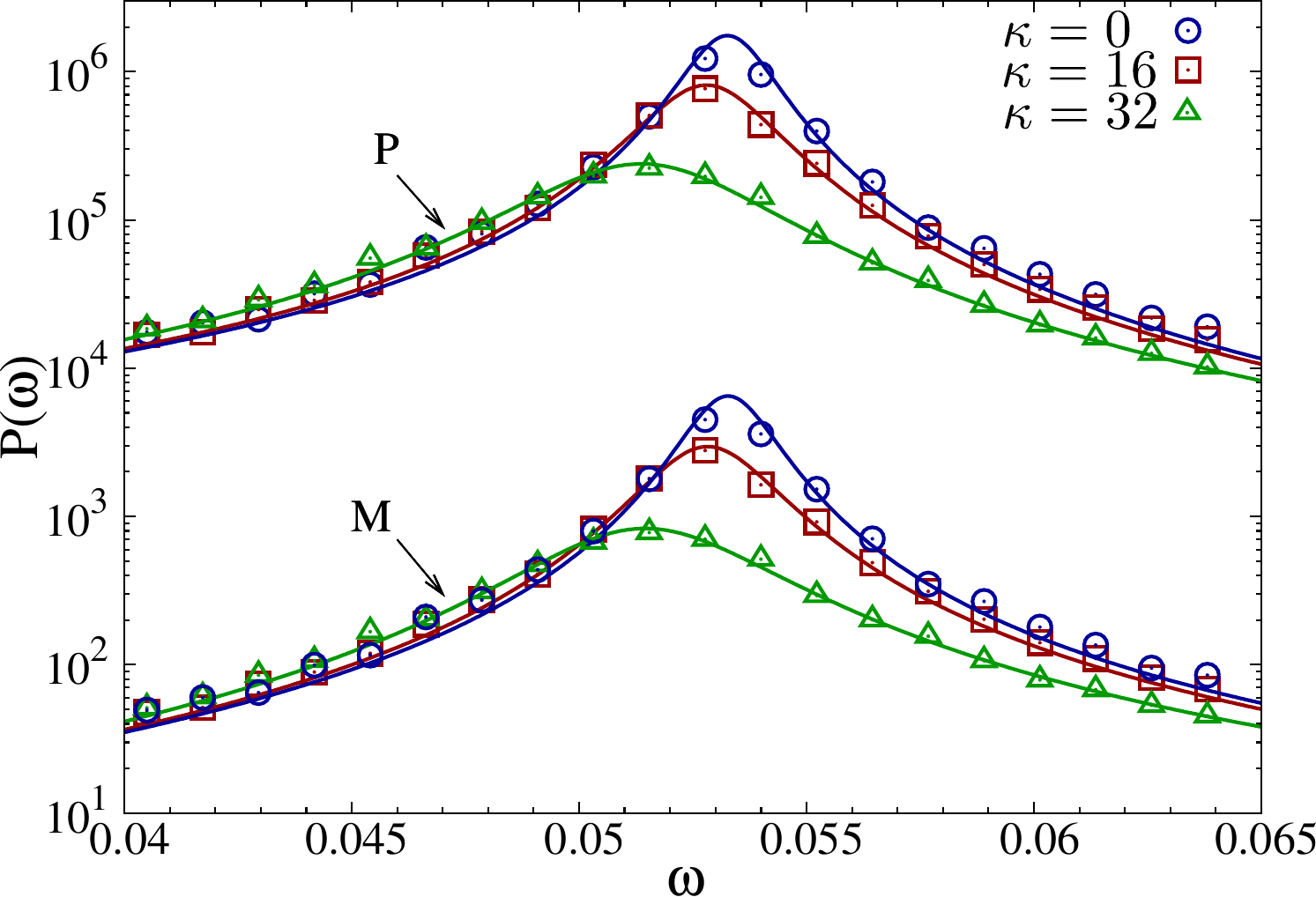}}
\vspace{0.5em}
\caption{(Colour on-line) Power spectra of quasi-cycles in the gene regulatory model \cite{monk,gene.2003,barrio} with uniformly distributed delays over the interval $[18.7-\kappa/2, 18.7+\kappa/2]$ minutes. Lines are theoretical predictions within the LNA, markers from simulations using a modified next-reaction method \cite{cai} and averaged over $700$ realisations. Parameters are $\alpha_M = \alpha_P = 1$, $\mu_P = \mu_M = 0.03$ (all with units min$^{-1}$), $P_0 = 10$, $h = 4.1$, $N = 5000$. }
\label{fig:gene}
\end{figure}
The first two interactions correspond to degradation of mRNA and protein, respectively, the constant model parameters $\mu_M$ and $\mu_P$ describe their degradation rates. The third interaction describes the translation of mRNA into protein. Finally, the fourth interaction represents the transcription process, within the model effectively the production of mRNA. This process is suppressed by the presence of protein molecules, as reflected by the Hill function, $g(n_P)=\alpha_M \left[1+[n_P/(P_0N)]^h\right]^{-1}$, where $h$ and $P_0$ are constants. The double arrow indicates a delay reaction. In this particular model the reaction has no effect on particle numbers at the time $t$ it is triggered, but only at a later time $t+\tau$, where $\tau$ is a distributed delay time drawn from $K(\tau)$.  The reaction rate depends on the number of proteins at the earlier time, $n_P(t)$.  Earlier works \cite{barrio,galla} focus on the case in which $K(\cdot)$ is a $\delta$-distribution, and exclude distributed delays.  
Applying our general result above (see Appendix~\ref{appendix:gene} for details) we find
\BE
\dot x_M(t)&=& \alpha_M \int_{0}^\infty d\tau ~  K(\tau)f[x_P(t-\tau)] \nonumber \\
&& - \mu_M x_M(t)+N^{-1/2}\eta_M(t), \nonumber \\
\dot x_P(t) &=& \alpha_P x_M(t) - \mu_P x_P(t)+N^{-1/2}\eta_P(t), \label{eq:gene}
\EE
where $f[x_P(t)]=[1+(x_P(t)/P_0)^h]^{-1}$, and
\BE
\avg{\eta_M(t) \eta_M(t')} &=& \bigg[\alpha_M \int_{0}^\infty d\tau ~  K(\tau)f[x_P(t-\tau)]\nonumber \\
&& + \mu_M x_M(t)\bigg] \delta(t-t'), \nonumber \\
\avg{\eta_P(t) \eta_P(t')} &=& \left[\alpha_P x_M(t) + \mu_P x_P(t)\right] \delta(t-t'), \nonumber \\
\avg{\eta_M(t) \eta_P(t')} &=& 0.
\EE
The Gaussian noise components $\eta_M, \eta_P$ have no correlations in time, as expected for a dynamics in which each reaction changes particle numbers only at one single time. A more complex case will be studied below. In the deterministic limit, $N\to\infty$, Eqs. (\ref{eq:gene}), are typically found to have a fixed point, ($x_M^*,x_P^*$). A systematic expansion, $x_M=x_M^*+N^{-1/2}\xi_M$, and similar for $x_P$, then leads to the LNA: a pair of linear SDEs for the fluctuation variables $\xi_M$ and $\xi_P$. A straightforward calculation following the lines of \cite{mckane} then allows one to compute the power spectra of noise-induced cycles, $P_M(\omega) = \avg{|\widetilde\xi_M(\omega)|^2}$, and similarly for the protein (see  Appendix~\ref{appendix:gene_ps}). Results for a uniform distribution of delay times are shown in Fig. \ref{fig:gene} and are confirmed convincingly in numerical simulations. In the LNA the stationary distribution for $\xi_M$ and $\xi_P$ can be derived as well, see Appendix~\ref{appendix:simulate} for further results and comparison against simulations.

\begin{figure}[t!!]
\centerline{\includegraphics[angle=270, width=0.4\textwidth]{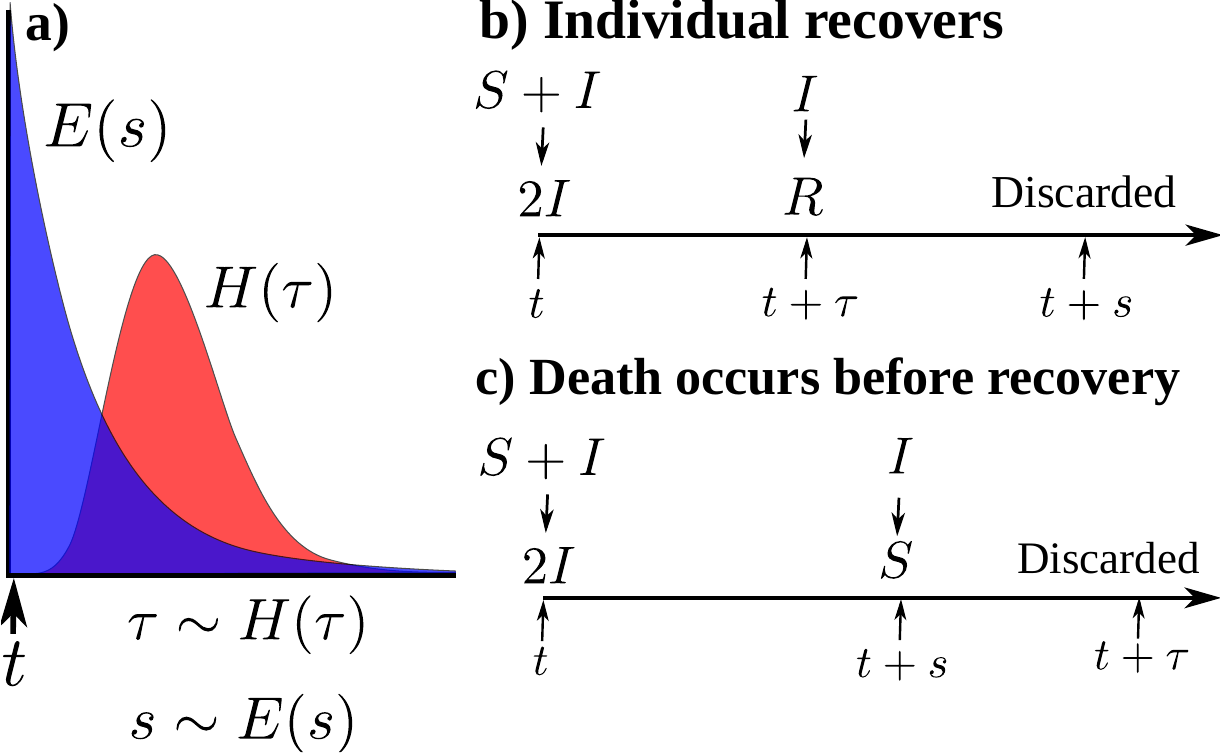}}
\vspace{0.5em}
\caption{(Colour on-line) Possible sequences of events when a reaction with delayed recovery is triggered. A time-to-death, $s$, and a time-to-recovery, $\tau$, are drawn from the appropriate distributions (panel a)). Depending on the outcome recovery or death may occur (panels b) and c) respectively), the remaining event is discarded.}
\label{fig:illustrate}
\end{figure}

{\em Application to a model of epidemic spread with delayed recovery.} We consider a variant of the susceptible-infective-recovered (SIR) model with birth and death \cite{andyalan}. The model describes a population of $N$ individuals, each of which can be in one of three states, $S$, $I$ or $R$. Infection occurs via the process $S+I\overset{\beta}\longrightarrow 2I$, and the newly infected individual may recover ($I\rightarrow R$) at a later time, where the delay is drawn from a distribution $H(\cdot)$. All individuals are subject to a birth-death process, occurring with rate $\mu$, and in which an individual dies and is immediately replaced by an individual of type $S$. This is a commonly used simplification, ensuring a constant population size \cite{andyalan}. This setup implies that a newly infected individual may die and be replaced by an individual of type $S$ before its designated recovery time is reached. This is illustrated in Fig. \ref{fig:illustrate}. Assume an infection occurs at time $t$. One may 
think of the subsequent dynamics as follows: at the time of infection, a designated time-to-recovery, $\tau$, is drawn from $H(\cdot)$. At the same time a designated time-to-removal, $s$, is drawn from an exponential distribution, $E(s)=\mu e^{-\mu s}$. There are then two possible subsequent courses of events: (i) If $\tau<s$ recovery occurs before death, the recovery process completes at time $t+\tau$, and the infective individual is replaced by an individual of type $R$. The death event is discarded. The probability for case (i) to occur is $\chi=\int_0^\infty d\tau H(\tau)\int_\tau^\infty ds~ E(s)$. Conditioned on this sequence of events, i.e. if  recovery occurring before death is a given, the time-to-recovery follows the distribution $K(\tau)= \chi^{-1}H(\tau)\int_{\tau}^\infty ds~ E(s)$. Case (ii) describes the opposite situation, $s<\tau$, occurring with probability $1-\chi$. In this case the newly infected individual dies before the designated time of recovery, and we have a reaction of type $I\to S$ 
at time $t+s$.  The conditional time-to-removal, given that case (ii) is realised, is $Q(s) = (1-\chi)^{-1}E(s) \int_{s}^\infty d\tau~ H(\tau)$.
We can summarise the reaction scheme as follows
\BE
R &\stackrel{\mu}{\longrightarrow}& S, \nonumber \\
S+I&\stackrel{\chi\beta}{\longrightarrow}& 2I; I\stackrel{K(\tau)}{\Longrightarrow} R, \nonumber \\
S+I&\stackrel{(1-\chi)\beta}{\longrightarrow}& 2I; I\stackrel{Q(s)}{\Longrightarrow} S.  
\EE
The notation for the second reaction channel, occurring with rate $T_2(\bn)=\beta\chi n_S n_I/N$, indicates that one particle of type $S$ is converted into an $I$ at the time the reaction is triggered, and that an individual of type $I$ is converted to $R$ at a later time $t+\tau$, where $\tau$ is drawn from the distribution $K(\cdot)$. Similarly, the third reaction channel fires with rate $T_3(\bn)=\beta(1-\chi) n_S n_I/N$, and results in an event $S+I\to 2I$ at the time the reaction is triggered, and then in an event of type $I\to S$ at a later time $t+s$, where $s$ is drawn from the distribution $Q(\cdot)$.

Applying the general result above we find (with $S=n_S/N, I=n_I/N$),

\BE
 \dot S(t) &=&-\beta S(t)I(t) +\mu(1-S(t)-I(t)) \nonumber \\
&&\hspace{-2em}+\beta (1-\chi)\int_{-\infty}^t dt' ~ Q(t-t') S(t')I(t')+N^{-1/2}\eta_S(t), \nonumber \\
\dot I(t) &=& \beta S(t)I(t) -\beta \int_{-\infty}^t dt'  S(t')I(t') \nonumber \\
&& \hspace{-2em}\times \bigg[\chi K(t-t')+(1-\chi)Q(t-t')\bigg] +N^{-1/2} \eta_I(t).\label{eq:sir}
\EE 
\begin{figure}[t!!!!]
\centerline{\includegraphics[angle=270, width=0.5\textwidth]{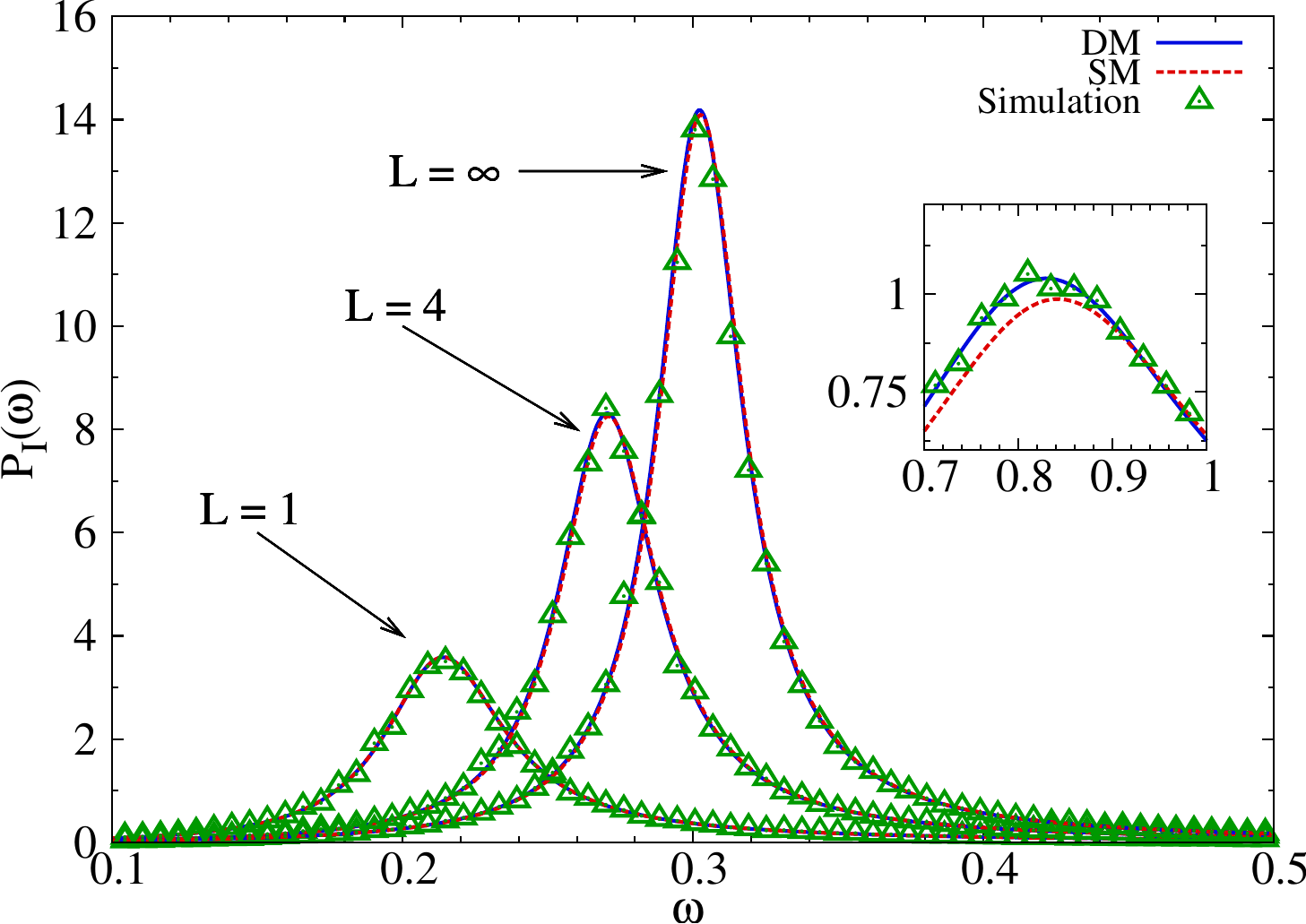}}
\caption{(Colour on-line) Power spectra $P_I(\omega)$ for the SIR model with delayed recovery. Lines show results from the LNA for the staged model (SM), see \cite{andyalan,remark2}, and for the delay model (DM). Markers are from simulations (SM for $L=1$ and $L=4$, DM for $L=\infty$), averaged over $800$ independent runs. Model parameters are $\beta=10.56 , \mu= 4.81\cdot 10^{-3}, \gamma = 1$. System size is $N=10^6$. Inset: Results for $L=4$ and $\mu= 4.81\cdot 10^{-2}$. }
\label{fig:sir}
\end{figure}

Unlike in the above model of gene regulation, the noise is now correlated in time. Expressions for the correlation matrix are lengthy and reported in Appendix~\ref{appendix:sir}.

Recent theoretical work has studied SIR-models in which individuals progress through a series of $L$ infectious `stages', $I_1\to I_2\to \dots \to I_L \to R$, at rate $\gamma L$ before they recover (or die along the way) \cite{andyalan}. In our formalism this is equivalent to a model in which $H(\cdot)$ is a $\Gamma$-distribution, $H(\tau) = \frac{(\gamma L)^L}{\Gamma(L)}\tau^{L-1}e^{-\gamma L \tau}$.  To make contact with the results of \cite{andyalan} we use Eqs. (\ref{eq:sir}) to compute the power spectra of noise-driven quasi-cycles about the fixed point of the deterministic limiting dynamics (see Appendix~\ref{appendix:sir_ps}). Results from the theory and from simulations are shown in Fig.~\ref{fig:sir}. We find that simulations of the staged model are less costly than those of the delay model. However the analytical calculation of the results in Fig. \ref{fig:sir} is more demanding in the staged model, as it involves a larger number of particle types. For sufficiently small values of the death rate, $\mu$, there is no 
noticeable difference between the predictions of our approach and the result of \cite{andyalan} (see main panel). This latter result is based on an expansion in $\mu$ and deviates from simulations when the small-$\mu$ approximation is not justified. Our theory does not rely on such approximations, and describes simulation results accurately in such cases (see inset of  Fig. \ref{fig:sir}). The staged model is limited to $\Gamma$-distributed recovery times, whereas our approach is more general and applies to other delay kernels suggested in the literature \cite{kernels}. Additional results can be found in Appendix~\ref{appendix:sir_ps}.

\emph{Conclusions.}
We have presented a comprehensive approach to the LNA for stochastic dynamics with distributed delays. Our calculation is based on a generating functional, rather than a master equation. We focus on probabilities to observe entire paths of the dynamics. This makes the approach suitable for non-Markovian systems, and we are able to derive general expressions for the Gaussian approximation of a broad class of processes with distributed delays. The resulting nonlinear chemical Langevin equation cannot normally be solved analytically, but it can be used for efficient simulations. Further analytical progress can be made in the linear-noise approximation. The validity of our results is demonstrated through the computation of power spectra of noise-driven cycles in delay models of gene regulation and of epidemic spread. We expect that the general expressions we have derived will be of use for studies of a variety of phenomena in the biological and physical sciences, and indeed in other areas where individual-based 
models with delayed interactions are relevant. 

{\em Acknowledgements.} TB acknowledges support from the EPSRC. TG is supported by RCUK (reference EP/E500048/1).

\onecolumngrid

\appendix

\section{System-size expansion without a master equation: Markovian systems}
\subsection{General remarks}
In this section we describe an alternative method with which to obtain results of the well-established expansion methods for Markovian systems. These expansions are usually carried out starting from a master equation describing the underlying stochastic process, and then following the procedure originally first proposed by van Kampen \cite{vk}, or alternatively the steps of the Kramers-Moyal expansion \cite{gardiner}. We here choose a different approach, and use a moment-generating function as the starting point. This technique is originally due to Martin, Siggia, Rose, Janssen and De Dominicis (MSRJD) \cite{msr}.

The central object underpinning the generating function approach is the probability measure of possible dynamical paths of the system, and not the probability of finding the system in a given state at a specific time. We will first discretise time in steps of duration $\Delta$, and keep $\Delta$ finite while we perform the average over paths and while we carry out the expansion in the system size at the level of the generating function. The limits of large $N$ and $\Delta\to 0$ are therefore decoupled, see also \cite{coolen} for a similar approach in a different context. It is only at the end that we take the the time step to zero, $\Delta\to 0$. This then results in a generating functional for a Gaussian process in continuous time, equivalent to the Gaussian dynamics one would have obtained from a direct Kramers-Moyal expansion of the underlying master equation or from Kurtz' theorem \cite{kurtz}. This multiplicative Gaussian process can then be linearised straightforwardly. For Markovian processes our 
method reproduces the known results of the linear-noise approximation (LNA). 

The generating function approach we take here is related to, but different from the path-integral formalism introduced by Doi and Peliti \cite{doipeliti}. The latter starts from a master equation, which is not required in the MSRJD-formalism. For Markovian systems the two approaches can be shown to lead to the same results, see for example \cite{lefevre} for a discussion. We also note that the outcome of van Kampen's system-size expansion for Markovian systems can be obtained in the Doi-Peliti formalism \cite{goldenfeld}.

The true strength of the generating function method becomes more transparent when we turn to non-Markovian delay dynamics. In such cases there is no closed master equation, and so we feel the generating function approach, which focuses on entire paths of the dynamics, is the most transparent technique to deal with delay dynamics.

\subsection{A Markovian example - model definitions}
We first illustrate the method using a specific example of a Markovian system, defined by the following set of reactions
 \BE
 \emptyset &\stackrel{1}{\longrightarrow}& A, \nonumber \\
 A+B &\stackrel{a}{\longrightarrow}& 3B, \nonumber \\
 B&\stackrel{b}{\longrightarrow}&\emptyset.
 \EE
This notation indicates that the three reactions occur with rates
 \be\label{eq:rates}
 T_1=N, ~~ T_2=a \frac{n m}{N}, ~~ T_3=b m.
 \ee 
 We have here written $n$ for the number of particles of type $A$ in the system, and $m$ for the number of $B$-particles. The quantity $N$ is a model parameter, and sets the scale of the system size. In a continuous-time setting reactions occur as exponential processes, the quantity $T_i[n(t),m(t)]dt$ represents the probability for a reaction of type $i$ to occur in the time interval $(t,t+dt)$.
 
\subsection{Discretisation of time and the generating function}
\label{appendix:disc_time}
In order to define the generating function we will discretise time into time steps of duration $\Delta$. This requires further qualification of the above reaction rates. We will be looking at a discrete time process, with paths $\{n_t,m_t\}=\dots, (n_{t-\Delta},m_{t-\Delta}), (n_t,m_t), (n_{t+\Delta},m_{t+\Delta}), \dots$. 

The dynamics of this system is determined by how frequently each of the three reactions above fire. We will write $k_{i,t}$ for the number of reactions of type $i=1, 2, 3$, which occur between time $t$ and time $t+\Delta$. Consistent with the above reaction rates we will assume that $k_{i,t}$ is a Poissonian random variable with mean $T_i(n_t,m_t)\Delta$. This amounts to a discretisation of the above continuous-time exponential process, similar to what is considered in the context of the algorithm with so-called $\tau$-leaping \cite{gillespie.2001}. We note that, due to the Poissonian character of the number of reactions that fire, particle numbers, $n_t, m_t$, can in principle go negative with non-zero probability in this discrete setup. We will however take the limit $\Delta\to 0$ at the end of the calculation. In this limit  unphysical negative particle numbers do not occur.

The starting point of our calculation is the Martin-Siggia-Rose-Janssen-De Dominicis (MSRJD) moment-generating function\footnote{We will use the term `generating function' for systems in discrete time, and `generating functional' when time is continuous.} \cite{msr}, see also \cite{aron, andreanov, altland, lefevre},
\BE
Z[\boldpsi,\boldphi]&=&\sum_{\bk}  \int D\bx D\by  ~{\cal P}(\bk)\left[\prod_t \delta\left(x_{t+\Delta}-x_t-\frac{k_{1,t}}{N}+\frac{k_{2,t}}{N}\right)\delta\left(y_{t+\Delta}-y_t-\frac{2k_{2,t}}{N}+\frac{k_{3,t}}{N}\right)\right]\nonumber \\
&&~~~~~~~~~~~~~~~~~~~~~~~~~~~~\times\exp\left(i\Delta\sum_t\left[\psi_tx_t+\varphi_t y_t\right]\right). \label{eq:appendixgf}
\EE
We have here introduced the variables $x_t=\frac{n_t}{N}$ and $y_t=\frac{m_t}{N}$, and we have written $D\bx=\prod_t dx_t$, and similar for $D\by$. We discuss the absence of any Jacobian determinants in the generating function below in the context of the more general model with delay (see Sec. \ref{sec:delay}). It should be noted that $t$ comes in integer multiples of the time step $\Delta$, i.e. we have $t=\ell\Delta$, where $\ell=\dots, -2,-1,0,1,2,\dots$. Our notation always implies this convention, it is important to keep this in mind when it comes to objects such as $\sum_t \psi_tx_t$, a short-hand for $\sum_{\ell\in\mathbb{Z}}\psi_{\ell\Delta}x_{\ell\Delta}$.

The quantity ${\cal P}(\bk)$ describes the joint probability distribution of the $\{k_{i,t}\}$, we note that the statistics of $k_{i,t}$ depends on $x_t$ and $y_t$. This dependence has been suppressed in our notation. The source terms $\boldpsi=\{\psi_t\}$ and $\boldphi=\{\varphi_t\}$ finally have been introduced as per normal procedure \cite{msr}. Taking derivatives with respect to these source terms generates moments of the variables $\{x_t\}$ and $\{y_t\}$, for example one has $\avg{x_t}=(i\Delta)^{-1}\left.\frac{\delta Z[\boldpsi,\boldphi]}{\delta\psi_t}\right|_{\boldpsi=\boldphi=0}$, where $\avg{\cdots}$ denotes the average  over all possible paths of the system,
\be
\avg{f[\bx,\by]}=\sum_{\bk}  \int D\bx D\by~ {\cal P}(\bk)\left[\prod_t \delta\left(x_{t+\Delta}-x_t-\frac{k_{1,t}}{N}+\frac{k_{2,t}}{N}\right)\delta\left(y_{t+\Delta}-y_t-\frac{2k_{2,t}}{N}+\frac{k_{3,t}}{N}\right)\right] f[\bx,\by].
\ee

\subsection{Average over trajectories and system-size expansion}
Starting from Eq. (\ref{eq:appendixgf}) we first write the delta-functions in the integrand in their exponential representation and obtain
\BE
Z[\boldpsi,\boldphi]&=&\sum_{\bk} \int\left[\prod_t\frac{dx_t d\widehat x_t}{2\pi}\frac{dy_td\hy_t}{2\pi}\right]{\cal P}(\bk) \exp\left(i\Delta \sum_t\left\{\psi_tx_t+\varphi_t y_t\right\}\right)\nonumber \\
&&\hspace{-3em} \times \exp\left(i\sum_t \left\{\hx_t\left[x_{t+\Delta}-x_t-\frac{k_{1,t}}{N}+\frac{k_{2,t}}{N}\right]+\hy_t\left[y_{t+\Delta}-y_t-\frac{2k_{2,t}}{N}+\frac{k_{3,t}}{N}\right]\right\}\right).
\EE
 
Next we collect terms and perform the average over the $\{k_{i,t}\}$. The relevant terms are
\be
\sum_{\bk} {\cal P}(\bk)  \exp\left(\frac{i}{N}\sum_t (-k_{1,t}\hx_t+k_{2,t}(\hx_t-2\hy_t)+k_{3,t}\hy_t\right).
\ee
The distribution of $\{k_{i,t}\}$ factorises iteratively according to ${\cal P}(\bk)=\prod_i\prod_t P_{i,t}(k_{i,t}|\{k_{j,t'}\}_{t'<t})$. I.e. in each realisation the statistics of the $\{k_{i,t}\}$ at a given time $t$ depend on the random numbers drawn at earlier times, $t'<t$. The individual factors can be written equivalently as $P_{i,t}(\,\cdot\,|\{k_{j,t'}\}_{t'<t}) = P_{i,t}(\,\cdot\,|x_t,y_t)$, and are Poissonian distributions with parameters $\lambda_{i,t}=T_i(n_t,m_t)\Delta$. 

As a specimen and to demonstrate the procedure, we now carry out the averaging over one of the $k_{i,t}$, specifically we choose $k_{2,t}$ for a fixed time step $t$. We have
 \BE
 \sum_{k_{2,t}}P_{2,t}(k_{2,t}|\{k_{j,t'}\}_{t'<t})\exp\left(i\frac{\hx_t-2\hy_t}{N}k_{2,t}\right) &=& \sum_{k_{2,t}}e^{-\lambda_{2,t}}\frac{\lambda_{2,t}^{k_{2,t}}}{k_{2,t}!}\exp\left(i\frac{\hx_t-2\hy_t}{N}k_{2,t}\right) \nonumber \\
 &=& e^{-\lambda_{2,t}}\sum_{k_{2,t}}\frac{\left[\lambda_{2,t}e^{i\frac{\hx_t-2\hy_t}{N}}\right]^{k_{2,t}}}{k_{2,t}!}  \nonumber \\
&=&\exp\left(-\lambda_{2,t}+\lambda_{2,t} e^{i\frac{\hx_t-2\hy_t}{N}}\right). 
\EE
This step is carried out for all reactions $i$, and it is iterated for all times. We note that the $\lambda_{i,t}$ at a given time $t$ depend on the particle numbers, $n_t, m_t$, at that time. The next step is to expand the above expression in powers of $N^{-1}$. We obtain
\BE
\exp\left(-\lambda_{2,t}+\lambda_{2,t} e^{i\frac{\hx_t-2\hy_t}{N}}\right)=\exp\left(i\lambda_{2,t}\frac{\hx_t-2\hy_t}{N}-\frac{1}{2}\lambda_{2,t}\frac{(\hx_t-2\hy_t)^2}{N^2}+\lambda_{2,t}\times {\cal O}(N^{-3})\right).
\EE
Keeping only terms of leading and sub-leading orders we arrive at the following expression for the generating function 
\BE
Z[\boldpsi,\boldphi]&=&\int\left[\prod_t\frac{dx_t d\widehat x_t}{2\pi}\frac{dy_td\hy_t}{2\pi}\right] \exp\left(i\sum_t\left\{ \hx_t(x_{t+1}-x_t)+\hy_t(y_{t+1}-y_t)\right\}\right)\exp\left(i\Delta \sum_t\hx_t(-r_{1,t}+r_{2,t})\right)\nonumber \\
&&\times\exp\Bigg(i\Delta \sum_t\hy_t(-2r_{2,t}+r_{3,t})-\frac{\Delta}{2N}\sum_t\left\{\hx_t^2(r_{1,t}+r_{2,t})+\hy_t^2(4r_{2,t}+r_{3,t})-4\hx_t\hy_tr_{2,t}\right\} \nonumber \\
&&~~~~~~~~~~~~~~~+\mbox{higher order terms}\Bigg)\times \exp\left(i\Delta \sum_t\left[\psi_tx_t+\varphi_t y_t\right]\right).\label{eq:gffinal}
\EE
We have here introduced the quantities $r_{i,t}$. These are shorthands for $r_i(x_t,y_t)$, where $\lambda_{i,t}=T_i(n_t,m_t)\Delta\equiv N\Delta r_i(x_t,y_t)$. The rates $T_i$ are of order $N$, so the $r_{i,t}$ are of order $N^0$.

\medskip
Dropping all terms beyond leading and sub-leading order expression (\ref{eq:gffinal}) can be re-written as

\BE
Z[\boldpsi,\boldphi]&=&\int\left[\prod_t\frac{dx_t d\widehat x_t}{2\pi}\frac{dy_td\hy_t}{2\pi}\right] \exp\left(i\Delta \sum_t \hx_t\left(\frac{x_{t+1}-x_t}{\Delta}-r_{1,t}+r_{2,t}\right)\right)\nonumber \\
&&\exp\left(i\Delta \sum_t\hy_t\left(\frac{y_{t+1}-y_t}{\Delta}-2r_{2,t}+r_{3,t}\right)\right) \exp\left(i\Delta \sum_t\left[\psi_tx_t+\varphi_t y_t\right]\right)\nonumber \\
&& \exp\left(-\frac{\Delta^2}{2N}\sum_{t,t'}\left\{\hx_t \cB_{x,x,t,t'} \hx_{t'} + \hy_t \cB_{y,y,t,t'} \hy_{t'}+ 2\hx_t \cB_{x,y,t,t'} \hy_{t'}\right\}+\dots\right),
 \EE
with
\BE
\cB_{x,x,t,t'}&=&\left(r_{1,t}+r_{2,t}\right) \frac{\delta_{t,t'}}{\Delta},\nonumber \\
\cB_{y,y,t,t'}&=&\left(4r_{2,t}+r_{3,t}\right)\frac{\delta_{t,t'}}{\Delta}, \nonumber \\
\cB_{x,y,t,t'}&=&-2r_{2,t}\frac{\delta_{t,t'}}{\Delta}.
\EE
\subsection{Continuous-time limit}
Taking the limit $\Delta\to 0$ we find to sub-leading order
\BE
Z[\boldpsi,\boldphi]&=&\int\D\bx\D\widehat\bx \D\by \D\widehat\by  \exp\left(i\int dt ~ \hx(t)\left[\dot x(t)-r_{1}[x(t),y(t)]+r_{2}[x(t),y(t)]\right]\right)\nonumber \\
&&\exp\left(i\int dt ~ \hy(t)\left[\dot y(t)-2r_{2}[x(t),y(t)]+r_{3}[x(t),y(t)]\right]\right)\nonumber \\
&&\times \exp\left(-\frac{1}{2N}\int dt~dt'~ \left\{\hx(t) B_{x,x}(t,t') \hx(t') + \hy(t) B_{y,y}(t,t') \hy(t')+ 2\hx(t) B_{x,y}(t,t') \hy(t')\right\}+\dots\right)\nonumber \\
 &&\times\exp\left(i\int dt ~\left[\psi(t)x(t)+\varphi(t) y(t)\right]\right),\label{eq:gfcont}
 \EE
where the (path-) integral $\int \D\bx \dots$ runs over all continuous-time paths $\{x(t)\}$, and similarly for $\int\D\by\dots$ and the auxiliary variables \cite{msr}. We have also introduced
\BE
B_{x,x}(t,t')&=&\big\{r_{1}[x(t),y(t)]+r_{2}[x(t),y(t)]\big\}\delta(t-t'), \nonumber \\
B_{y,y}(t,t')&=&\big\{4r_{2}[x(t),y(t)]+r_{3}[x(t),y(t)]\big\}\delta(t-t'), \nonumber \\
B_{x,y}(t,t')&=&-2r_{2}[x(t),y(t)]\delta(t-t').
\EE
We have here used the correspondence $\Delta^{-1}\delta_{t,t'}\leftrightarrow \delta(t-t')$ between the Kronecker-$\delta$ for discrete arguments, and the Dirac $\delta$-function for continuous arguments. This correspondence can easily be verified using the correspondence $\Delta\sum_t f_t\leftrightarrow \int dt f(t)$, as well as $\int dt' \delta(t-t') f(t')=f(t)$, and the observation that $\sum_{t'} \delta_{t,t'} f_{t'}=\Delta\sum_{t'} (\Delta^{-1}\delta_{t,t'})f_{t'}=f_t$.

The expression in Eq. (\ref{eq:gfcont}) is recognised as the generating functional of the following dynamics 
\BE
\dot x(t)&=&r_1[x(t),y(t)]-r_2[x(t),y(t)]+\frac{1}{\sqrt{N}}\eta(t), \nonumber \\
\dot y(t)&=&2r_2[x(t),y(t)]-r_3[x(t),y(t)]+\frac{1}{\sqrt{N}}\zeta(t), \label{eq:m1}
\EE 
with white Gaussian noise variables $\eta(t)$ and $\zeta(t)$ (both of mean zero), and with correlations
\BE
\avg{\eta(t)\eta(t')}&=&\left\{r_1[x(t),y(t)]+r_2[x(t),y(t)]\right\} \delta(t-t'),\nonumber \\
\avg{\zeta(t)\zeta(t')}&=&\left\{4r_2[x(t),y(t)]+r_3[x(t),y(t)]\right\}\delta(t-t'),\nonumber \\
\avg{\eta(t)\zeta(t')}&=&-2r_2[x(t),y(t)]\delta(t-t').\label{eq:m2}
\EE
This is the result one would have obtained from a direct Kramers-Moyal expansion of the master equation of the system \cite{gardiner} or by applying Kurtz' theorem \cite{kurtz}. For later convenience we write $F_1(x,y)=r_1(x,y)-r_2(x,y)$ and $F_2(x,y)=2r_2(x,y)-r_3(x,y)$. In our specific example one has $r_1[x(t),y(t)]=1$, $r_2[x(t),y(t)]=ax(t)y(t)$ and $r_3[x(t),y(t)]=y(t)$.
\subsection{Linear-noise approximation}
The result obtained in the previous section, Eqs. (\ref{eq:m1}, \ref{eq:m2}), describes a process with multiplicative noise, and is a slightly stronger result than the LNA. The latter is obtained by writing $x(t)=x^\infty(t)+N^{-1/2}\xi_1(t)$ and $y(t)=y^\infty(t)+N^{-1/2}\xi_2(t)$, where the deterministic trajectory $x^\infty(t), y^\infty(t)$ is the solution of $\dot x^\infty=F_1(x^\infty,y^\infty), \dot y^\infty=F_2(x^\infty,y^\infty)$. The superscript `$\infty$' indicates that this deterministic trajectory is obtained in the thermodynamic limit, $N\to \infty$.

Inserting into Eqs. (\ref{eq:m1}), and expanding in powers of $N^{-1/2}$, one then finds
\BE
\dot \xi_1 &=& \frac{\partial F_1(x^\infty,y^\infty)}{\partial x^\infty}\xi_1+\frac{\partial F_1(x^\infty,y^\infty)}{\partial y^\infty}\xi_2+ \eta_1, \nonumber \\
\dot \xi_2 &=& \frac{\partial F_2(x^\infty,y^\infty)}{\partial x^\infty}\xi_1+\frac{\partial F_2(x^\infty,y^\infty)}{\partial y^\infty}\xi_2+ \eta_2,
\EE
where
\BE
\avg{\eta_1(t)\eta_1(t')}&=&\left\{r_1[x^\infty(t),y^\infty(t)]+r_2[x^\infty(t),y^\infty(t)]\right\} \delta(t-t'),\nonumber \\
\avg{\eta_2(t)\eta_2(t')}&=&\left\{4r_2[x^\infty(t),y^\infty(t)]+r_3[x^\infty(t),y^\infty(t)]\right\}\delta(t-t'),\nonumber \\
\avg{\eta_1(t)\eta_2(t')}&=&-2r_2[x^\infty(t),y^\infty(t)]\delta(t-t'). 
\EE
The noise is now linear (additive).
\subsection{General theory for Markovian processes}
We now consider a general Markovian model with $S$ types of particles, $\alpha=1,\dots, S$ and $M$ reactions, $i=1,\dots,M$, again in discrete time. We will write $n_{\alpha,t}$ for number of particles of type $\alpha$ in the system at time $t$, and we introduce $\bx=(x_1,\dots,x_S)$, where $x_\alpha=n_\alpha/N$. We assume that the reactions occur with rates $T_i=\Delta N r_i$, where the $r_i$ is of order $N^0$, and may depend on the state of the system, $r_i=r_i(\bx)$. We will further assume that the occurrence of a reaction of type $i$ will result in a change of the number of particles of type $\alpha$ by $v_{i,\alpha}$, so that $n_{\alpha,t+\Delta}=n_{\alpha,t}+\sum_i k_{i,t}v_{i,\alpha}$ if $k_{i,t}$ reactions of type $i$ occur between time $t$ and time $t+\Delta$.

The generating function is now given by 
\BE
Z[\boldpsi]&=&\sum_{\bk} \int D\bx ~{\cal P}(\bk)  \left[\prod_{t,\alpha} \delta\left(x_{\alpha, t+\Delta}-x_{\alpha,t}-\sum_i \frac{k_{i,t}v_{i,\alpha}}{N}\right)\right]\exp\left(i\Delta\sum_{t,\alpha}\psi_{\alpha,t} x_{\alpha,t}\right) \nonumber \\
&=&\sum_{\bk} \int\left[\prod_{\alpha,t}\frac{dx_{\alpha,t} d\widehat x_{\alpha,t}}{2\pi}\right] {\cal P}(\bk)\exp\left(i\sum_{\alpha,t}\hx_{\alpha,t}\left(x_{\alpha, t+\Delta}-x_{\alpha,t}-\sum_i \frac{k_{i,t}v_{i,\alpha}}{N}\right)\right]\exp\left(i\Delta\sum_{t,\alpha}\psi_{\alpha,t} x_{\alpha,t}\right). \nonumber \\
\EE
The terms containing a given Poissonian variable $k_{i,t}$ are now
\BE
&&\sum_{k_{i,t}} P_{i,t}(k_{i,t}|\{k_{j,t'}\}_{t'<t})\exp\left(-\frac{i}{N}\sum_\alpha \hx_{\alpha,t}k_{i,t}v_{i,\alpha}\right)\nonumber \\
&=&\exp\left(-\frac{i}{N}\lambda_{i,t}\sum_\alpha \hx_{\alpha,t}v_{i,\alpha}-\frac{1}{2N^2}\lambda_{i,t}\left(\sum_\alpha \hx_{\alpha,t}v_{i,\alpha}\right)^2+\dots \right)\nonumber \\
&=&\exp\left(-\frac{i}{N}\lambda_{i,t}\sum_\alpha \hx_{\alpha,t}v_{i,\alpha}-\frac{1}{2N^2}\lambda_{i,t}\sum_{\alpha,\beta} \hx_{\alpha,t}v_{i,\alpha}v_{i,\beta}\hx_{\beta,t}+\dots \right).
\EE
As before we here work to sub-leading order. This leads to 
\BE
Z[\boldpsi]&=&\int\left[\prod_{\alpha,t}\frac{dx_{\alpha,t} d\widehat x_{\alpha,t}}{2\pi}\right] \exp\left(i\Delta \sum_{\alpha,t}\hx_{\alpha,t}\left(\frac{x_{\alpha, t+\Delta}-x_{\alpha,t}}{\Delta}-\sum_ir_{i,t}v_{i,\alpha}\right)\right)\nonumber \\
&&\times \exp\left(-\frac{\Delta^2}{2N}\sum_{t,t'} \sum_{\alpha,\beta} \hx_{\alpha,t}\cB_{\alpha,\beta,t,t'}\hx_{\beta,t'}+\dots \right)\exp\left(i\Delta\sum_{t,\alpha}\psi_{\alpha,t} x_{\alpha,t}\right).
\EE
We have here introduced
\be
\cB_{\alpha,\beta,t,t'}=\left(\sum_i r_{i,t}v_{i,\alpha}v_{i,\beta}\right)\frac{\delta_{t,t'}}{\Delta}.
\ee
In the limit $\Delta\to 0$ we find
\BE
Z[\boldpsi]&=&\int\D\bx\D\widehat\bx \exp\left(i\sum_{\alpha}\int dt ~ \hx_{\alpha}(t)\left(\dot x_\alpha(t)-\sum_i r_{i}[\bx(t)] v_{i,\alpha}\right)\right)\nonumber \\
&&\times \exp\left(-\frac{1}{2N}\int dt~ dt' \sum_{\alpha,\beta} \hx_{\alpha}(t)B_{\alpha,\beta}(t,t')\hx_{\beta}(t')+\dots \right)\exp\left(i\sum_\alpha\int dt ~ \psi_{\alpha}(t) x_{\alpha}(t)\right),\label{eq:this}
\EE
where 
\be
B_{\alpha,\beta}(t,t')=\left(\sum_i r_{i}[\bx(t)]v_{i,\alpha}v_{i,\beta}\right)\delta(t-t'). 
\ee
 While we have indicated discrete-time arguments as subscripts in the preceding sections (e.g. $x_{\alpha,t}$), we will use a time argument in brackets ($x_\alpha(t)$) for continuous times. Eq. (\ref{eq:this}) in turn is the generating function description of the dynamics
\be
\dot x_{\alpha}(t)= \sum_i r_i[\bx(t)]v_{i,\alpha}+N^{-1/2}\eta_{\alpha}(t),
\ee
where the $\{\eta_\alpha(t)\}$ describe Gaussian white noise of mean zero and with correlations
\be
\avg{\eta_{\alpha}(t)\eta_{\beta}(t')}=\left(\sum_i v_{i,\alpha}v_{i,\beta}r_{i}[\bx(t)]\right)\delta(t-t')\label{eq:corr2}
\ee
across particle types. Again this is the result one would have obtained from a Kramers-Moyal expansion \cite{gardiner}, or from a direct application of Kurtz' theorem \cite{kurtz}.

The linear noise approximation is obtained following the steps outlined above. One writes $\bx(t)=\bx^\infty(t)+N^{-1/2}\boldxi(t)$, and finds
\be
\dot \xi_\alpha(t)=\sum_\beta J_{\alpha,\beta}[\bx^\infty(t)]\xi_\beta(t)+\eta_\alpha(t)\label{eq:linn}
\ee
where $J_{\alpha,\beta}=\partial F_{\alpha}/\partial x_\beta$, and where the replacement $\bx(t)\to\bx^\infty(t)$ is implied in the noise correlator given in Eq. (\ref{eq:corr2}). This makes the noise additive. We have here written $F_\alpha(\bx)= \sum_i r_i(\bx)v_{i,\alpha}$. Eq. (\ref{eq:linn}) represents the result one would have obtained from a direct application of van Kampen's system-size expansion \cite{vk}.
\section{General theory for delay systems}\label{sec:delay}
\subsection{Model definitions}
We will next look at a discrete-time system with delay interactions. We will again assume that there are $S$ types of particles, $\alpha=1,\dots, S$, with particle numbers $n_{\alpha,t}$, and that there are $M$ reactions, $i=1,\dots,M$. As before we write $x_{\alpha,t}=n_{\alpha,t}/N$. In the continuous-time model reaction $i$ is taken to fire with rate $T_i=N r_i$, where $r_i=r_i(\bx)\sim{\cal O}(N^0)$.  The immediate change in particle numbers at this time will be denoted by $v_{i,\alpha}$ as before.

A delay reaction firing at time $t$ can affect particle numbers at later time steps. Throughout this paper we assume that any instance of a reaction will only induce changes of particle numbers at {\em at most one single later time} (this is not a severe restriction and will be the case for most applications). This time is a random variable, drawn from an underlying probability distribution. We will further restrict the discussion to models in which the distribution of delay times, $\tau$, of a given reaction does not depend on the state of the system when the reaction is triggered. Again this is not a severe constraint, and is usually fulfilled (generalisation of our theory to models in which this is not the case are possible). To be more specific we will assume that once a reaction of type $i$ is triggered at time $t$ an immediate change of particle numbers by $v_{i,\alpha}$ occurs ($\alpha=1,\dots,S$). Additionally a delay time $\tau>0$ is drawn from a distribution $K_i(\cdot)$, and a further change of 
particle numbers by $w_{i,\alpha}^\tau$ then occurs at time $t+\tau$. The appropriate normalisation of $K_i(\cdot)$ is $\int_{0}^\infty d\tau K_i(\tau)=1$. The rate with which a reaction of type $i$ with a delay time of precisely $\tau$ fires at time $t$ is hence $T_i[\bn(t)]K_i(\tau)$. The probability to see a reaction of type $i$ fire during the time interval from $t$ to $t+dt$ and with a delay time in the interval from $\tau$ to $\tau+d\tau$ is $T_i(\bn(t))K_i(\tau)dt d\tau$.

This change in particle numbers at the later time can in principle depend on the drawn delay time $\tau$, hence the superscript in $w_{i,\alpha}^\tau$, even though this is not usually the case in most applications. We allow for this possibility when we develop the theory, in the two applications we study in this paper the $w_{i,\alpha}^\tau$ are independent of $\tau$.

We also point out that our formalism allows one to include reactions without delayed effect. These will simply have $w_{i,\alpha}^\tau\equiv 0$. In absence of any reactions with delay  ($w_{i,\alpha}^\tau=0$ for all $i,\alpha,\tau$) we recover the Markovian case discussed in the previous section.

\subsection{Generating function approach}
\label{appendix:delay_generating}
In the discrete-time setting (with time step $\Delta$) the number of reactions of type $i$ triggered (with any delay time $\tau$) at time step $t$ will be a Poissonian random variable $k_{i,t}$ with mean $T_i\Delta=N\Delta r_i(\bx_t)$. The number of reactions of type $i$ with a delayed effect $\tau=\ell\Delta$ time steps later will be a Poissonian random variable $k_{i,t}^\tau$ with mean $(\Delta T_i)\times(\Delta K_i(\tau))$. The second factor, $\Delta\times K_i(\tau)$, is the probability that a delay time in the time interval $(\tau,\tau+\Delta)$ is drawn in the continuum model. We have $k_{i,t}=\sum_\tau k_{i,t}^\tau$. For later convenience we will introduce $\rho_{i,\tau}=\Delta \times K_i(\tau)$, so that $k_{i,t}^\tau$ becomes a Poissonian random variable with mean $\lambda_{i,t}^\tau=N\Delta r_i(\bx_t)\rho_{i,\tau}$. The variable $k_{i,t}$ follows a Poissonian distribution with parameter $\lambda_{i,t}=\sum_\tau \lambda_{i,t}^\tau=N\Delta r_i(\bx_t)$.

\medskip

The MSRJD generating function for this process is given by\footnote{We note that potential functional determinants in this expression are field-independent and can hence be discarded, see also \cite{andreanov, aron, lefevre, altland}. If we introduce the short-hand $X_{\alpha,t+\Delta}$ for the expression inside the delta-function in Eq. (\ref{eq:gf0}), then the relevant Jacobian is given by ${\cal J}_{\alpha,\beta,t,t'}=\partial X_{\alpha,t}/\partial x_{\beta,t'}$.  Keeping in mind that the statistics of $k_{i,t}^\tau$ only depend on the state $\bx_t$, but not on states, $\bx_{t'}$ at times $t'>t$, one then notes that ${\cal J}_{\alpha,\beta,t,t'}=0$ for $t'>t$, due to causality. One also has ${\cal J}_{\alpha,\beta,t,t}=\delta_{\alpha,\beta}$. The matrix ${\cal J}$ is triangular with respect to the indices $t$ and $t'$, with unit diagonal elements. This makes its determinant field-independent and equal to unity. }

\BE
Z[\boldpsi]&=&\sum_{\bk} \int D\bx ~ {\cal P}(\bk) \left[\prod_{t,\alpha} \delta\left(x_{\alpha, t+\Delta}-x_{\alpha,t}-\sum_i k_{i,t}\frac{v_{i,\alpha}}{N}-\sum_{i}\sum_{\tau\geq \Delta} \frac{k_{i,t-\tau}^\tau w_{i,\alpha}^\tau}{N}\right)\right]\nonumber \\
&&\times\exp\left(i\Delta\sum_{t,\alpha}\psi_{\alpha,t} x_{\alpha,t}\right). \label{eq:gf0}
\EE
In order to keep the notation sufficiently compact we will simply write $\sum_\tau \dots$ in the following instead of $\sum_{\tau\geq\Delta}\dots$. The constraint $\tau\geq\Delta$ is always implied. 

Given that $k_{i,t}=\sum_{\tau\geq \Delta}k_{i,t}^\tau$ this can be written as
\BE
Z[\boldpsi]&=&\sum_{\bk}  \int D\bx  {\cal P}(\bk)\left[\prod_{t,\alpha} \delta\left(x_{\alpha, t+\Delta}-x_{\alpha,t}-\sum_{i,\tau} k_{i,t}^\tau\frac{v_{i,\alpha}}{N}-\sum_{i,\tau} \frac{k_{i,t-\tau}^\tau w_{i,\alpha}^\tau}{N}\right)\right]\exp\left(i\Delta\sum_{t,\alpha}\psi_{\alpha,t} x_{\alpha,t}\right) 
\nonumber \\
&=&\sum_{\bk}\int \left[\prod_{\alpha,t}\frac{dx_{\alpha,t} d\widehat x_{\alpha,t}}{2\pi}\right] {\cal P}(\bk) \exp\left(i\sum_{\alpha,t}\hx_{\alpha,t}\left(x_{\alpha, t+\Delta}-x_{\alpha,t}-\sum_{i,\tau} \frac{k_{i,t}^\tau v_{i,\alpha}+k_{i,t-\tau}^\tau w_{i,\alpha}^\tau}{N}\right)\right)\nonumber \\
&&\times \exp\left(i\Delta\sum_{t,\alpha}\psi_{\alpha,t} x_{\alpha,t}\right).
\EE

The terms containing the Poissonian variables $\bk$ are given by
\BE
&&\left\langle \exp\left(-\frac{i}{N}\sum_{i,t,\tau} k_{i,t}^\tau \sum_\alpha \left\{ \hx_{\alpha,t}v_{i,\alpha}+\hx_{\alpha,t+\tau}w_{i,\alpha}^{\tau} \right\}\right)\right\rangle_{\bk}\nonumber \\
 &=&\exp\left(-\frac{i}{N}\sum_{t,\alpha} \hx_{\alpha,t}\sum_i\left\{\lambda_{i,t}v_{i,\alpha}+\sum_\tau \lambda_{i,t-\tau}^\tau w_{i,\alpha}^\tau\right\}\right) \exp\left(-\frac{1}{2N^2}\sum_{t,\tau} \sum_{\alpha,\beta} \left\{2\hx_{\alpha,t}\left[\sum_i\lambda_{i,t}^\tau v_{i,\alpha}w_{i,\beta}^\tau\right] \hx_{\beta,t+\tau}\right\}\right) \nonumber \\
 &&\times \exp\left(-\frac{1}{2N^2}\sum_{t}\sum_{\alpha,\beta} \left\{\hx_{\alpha,t}\sum_i\left[ \lambda_{i,t}v_{i,\alpha}v_{i,\beta}+\sum_\tau \lambda_{i,t-\tau}^\tau w_{i,\alpha}^\tau w_{i,\beta}^\tau \right] \hx_{\beta,t}\right\} +{\cal O}(N^{-3})\right).
 \EE
 We have here written $\avg{\dots}_\bk$ for the average over the $\{k_{i,t}^\tau\}$, where one again keeps in mind that the statistics of the $\{k_{i,t}^\tau\}$ depend on the dynamics at earlier times, i.e. on the $\{k_{j,t}^\tau\}_{t'<t}$, and the statistics of the $\{k_{i,t}^\tau\}$ are determined by the state of the system at time $t$, $\bx(t)$.
 Putting the pieces together, and replacing $\lambda_{i,t}=N\Delta r_i(\bx_t)$, and $\lambda_{i,t}^\tau=N\Delta r_i(\bx_t)\rho_{i,\tau}$, we have
 \BE
Z[\boldpsi]&=&\int \left[\prod_{\alpha,t}\frac{dx_{\alpha,t} d\widehat x_{\alpha,t}}{2\pi}\right] \exp\left(i\sum_{\alpha,t}\hx_{\alpha,t}\left(x_{\alpha, t+\Delta}-x_{\alpha,t}\right)\right)\exp\left(i\Delta\sum_{t,\alpha}\psi_{\alpha,t} x_{\alpha,t}\right) \nonumber \\
&&\times \exp\left(-i\Delta\sum_{t,\alpha} \hx_{\alpha,t}\sum_i \left\{r_i(\bx_t)v_{i,\alpha}+\sum_\tau r_i(\bx_{t-\tau}) \rho_{i,\tau} w_{i,\alpha}^\tau\right\}\right) \nonumber \\
 &&\times \exp\left(-\frac{\Delta}{2N}\sum_{t}\sum_{\alpha,\beta} \left\{\hx_{\alpha,t}\sum_i \left[ r_i(\bx_t)v_{i,\alpha}v_{i,\beta}+\sum_\tau r_i(\bx_{t-\tau})\rho_{i,\tau}w_{i,\alpha}^\tau w_{i,\beta}^\tau \right] \hx_{\beta,t}\right\}\right)\nonumber \\ 
&& \times \exp\left(-\frac{\Delta}{2N}\sum_{t,\tau} \sum_{\alpha,\beta} \left\{2\hx_{\alpha,t}\left[\sum_i  r_i(\bx_t) \rho_{i,\tau} v_{i,\alpha}w_{i,\beta}^\tau\right] \hx_{\beta,t+\tau}\right\}+\dots\right),
\EE
where the ellipsis (`$\dots$') in the last exponential again indicate that all terms beyond leading and sub-leading order in $N^{-1}$ have been dropped. This can be written as
 \BE
Z[\boldpsi]&=&\int \left[\prod_{\alpha,t}\frac{dx_{\alpha,t} d\widehat x_{\alpha,t}}{2\pi}\right] \exp\left(i\Delta \sum_{\alpha,t}\hx_{\alpha,t}\left(\frac{x_{\alpha, t+\Delta}-x_{\alpha,t}}{\Delta}-f_{\alpha,t}(\bx)\right)\right)\exp\left(i\Delta\sum_{t,\alpha}\psi_{\alpha,t} x_{\alpha,t}\right) \nonumber \\
&&\times\exp\left(-\frac{\Delta^2}{2}\sum_{t,t'}\sum_{\alpha,\beta} \hx_{\alpha,t} \cB_{\alpha,\beta,t,t'}(\bx)\hx_{\beta,t'}+\dots\right),
\EE
where
\BE
f_{\alpha,t}(\bx)&=&\sum_i \left\{r_i(\bx_t)v_{i,\alpha}+\sum_\tau r_i(\bx_{t-\tau})\rho_{i,\tau} w_{i,\alpha}^\tau\right\}, \nonumber \\
\cB_{\alpha,\beta,t,t'}(\bx)&=&\frac{1}{N}\left\{\frac{1}{\Delta}\delta_{t,t'}\sum_i \left[ r_i(\bx_t)v_{i,\alpha}v_{i,\beta}+\sum_\tau r_i[\bx_{t-\tau}]\rho_{i,\tau}w_{i,\alpha}^\tau w_{i,\beta}^\tau \right]\right.\nonumber \\
&&\left.+\sum_i \left[ r_i(\bx_{t}) \frac{\rho_{i,t'-t}}{\Delta} v_{i,\alpha}w_{i,\beta}^{(t'-t)}+ r_i(\bx_{t'}) \frac{\rho_{i,t-t'}}{\Delta} v_{i,\beta}w_{i,\alpha}^{(t-t')}\right] \right\}.
\EE
The quantity $f_{\alpha,t}(\bx)$ depends on the trajectory of the system up to time $t$. Similarly, $\cB_{\alpha,\beta,t,t'}(\bx)$ is determined by the trajectory up to the larger of the two times, $t$ and $t'$. The notation indicates the dependence on the trajectory and the explicit time dependence.
\subsection{Continuous-time limit}
Taking the limit $\Delta\to 0$ we find
 \BE
Z[\boldpsi]&=&\int \D\bx \D\widehat\bx \exp\left(i \sum_{\alpha}\int dt ~\hx_{\alpha}(x)\left[\dot x_{\alpha}(t)-F_{\alpha}(t,\bx)\right]\right)\exp\left(i\sum_{\alpha}\int dt ~ \psi_{\alpha}(t) x_{\alpha}(t)\right) \nonumber \\
&&\times\exp\left(-\frac{1}{2N}\int dt~ dt' ~\sum_{\alpha,\beta} \hx_{\alpha}(t) B_{\alpha,\beta}(t,t',\bx)\hx_{\beta}(t')+\dots\right),\label{eq:gfcontinuous}
\EE
where
\BE
F_{\alpha}(t,\bx)&=&\sum_i \left\{r_i[\bx(t)]v_{i,\alpha}+\int_0^\infty d\tau~ r_i[\bx(t-\tau)] K_i(\tau) w_{i,\alpha}^\tau\right\}, \nonumber \\
B_{\alpha,\beta}(t,t',\bx)&=&\left\{\delta(t-t')\sum_i \left[ r_i[\bx(t)]v_{i,\alpha}v_{i,\beta}+\int_0^\infty d\tau~ r_i[\bx(t-\tau)]K_i(\tau)w_{i,\alpha}^\tau w_{i,\beta}^\tau \right]\right.\nonumber \\
&&\left.+\sum_i \left[  r_i[\bx(t)] K_i(t'-t) v_{i,\alpha}w_{i,\beta}^{(t'-t)}+ r_i[\bx(t')] K_i(t-t') v_{i,\beta}w_{i,\alpha}^{(t-t')}\right] \right\}. 
\EE
The quantities $F_{\alpha}(t,\bx)$ and $B_{\alpha,\beta}(t,t',\bx)$ again depend on the trajectory and they have an explicit dependence on their time arguments.

Eq. (\ref{eq:gfcontinuous}) corresponds to the dynamics
\be\label{eq:dyn}
\dot x_\alpha=F_\alpha(t,\bx)+N^{-1/2}\eta_\alpha,
\ee
where 

\BE
\avg{\eta_\alpha(t)\eta_\beta(t')}&=&\delta(t-t')\sum_i \left[ r_i[\bx(t)]v_{i,\alpha}v_{i,\beta}+\int_0^\infty d\tau~ r_i[\bx(t-\tau)]K_i(\tau)w_{i,\alpha}^\tau w_{i,\beta}^\tau \right]\nonumber \\
&&+\sum_i \left[  r_i[\bx(t)] K_i(t'-t) v_{i,\alpha}w_{i,\beta}^{(t'-t)}+ r_i[\bx(t')] K_i(t-t') v_{i,\beta}w_{i,\alpha}^{(t-t')}\right].\label{eq:corr3}
\EE

\subsection{Linear-noise approximation}
\label{appendix:lna}
One proceeds again by decomposing $\bx(t)=\bx^\infty(t)+N^{-1/2}\boldxi(t)$, where $\bx^\infty$ is the deterministic trajectory, i.e. the solution of
\be
\dot x_\alpha^\infty=F_\alpha(t,\bx^\infty).
\ee
Substituting in Eq. (\ref{eq:dyn}) we find
\be
\dot \boldxi(t)=\int _{-\infty}^t dt'~ \underline{\underline{J}}(t-t',\bx^\infty) \boldxi(t')+\boldeta(t),
\ee
where we again imply the substitution $\bx\to\bx^\infty$ in the correlator of the noise, see Eq. (\ref{eq:corr3}). The Jacobian is defined by
\be
J_{\alpha,\beta}(t,t',\bx)=\frac{\delta F_\alpha(t,\bx)}{\delta x_\beta(t')},
\ee
and depends only on time differences, $t-t'$.
 \section{Model of gene regulation with delay interaction}
 \label{appendix:gene}
\subsection{Gaussian approximation}
The delay model of gene regulation discussed in the main paper (see also \cite{barrio, bratsun}) describes two types of particles, $\alpha=P, M$. We will write
\BE
\bx = \left(\begin{array}{c} x_P\\x_M\end{array}\right) .
\EE
 In the model there are four reactions, $i=1,\dots, 4$, one of which is a delay reaction. The reaction rates are
\BE
\br[\bx(t)] = \left( \begin{array}{c} \mu_M x_M \\ \mu_P x_P \\ \alpha_P x_M \\ \alpha_Mf[x_P(t)]\end{array} \right) .
\EE
with  $f[x_P(t)]=[1+(x_P(t)/P_0)^h]^{-1}$. 

The effect of the reactions at the time they are triggered are described by the stoichiometric coefficients

\BE
\underline{\underline v} = \left (\begin{array}{cc} 0&-1\\-1&0\\1&0\\0&0\end{array}\right),
\EE
where each row represents one of the reactions, and where the first column is the change in number of protein-molecules and the second column is the change in the number of mRNA molecules.

There is only one delay reaction in this model, $i = 4$. The corresponding distribution of delay times is $K_4(\tau) = K(\tau)$, and the delayed effect on particle numbers is given by $w_{4,M} = 1$. All other $w_{i,\alpha}$ vanish. 

Applying the general result discussed in the main paper [Eqs. (3, 4, 5)] we find
\BE\label{eq:kramers1}
F_{P}(t, \bx)&=&\alpha_P x_M(t) - \mu_P x_P(t), \nonumber \\
F_M(t,\bx) &=& \alpha_M \int_{-\infty}^t dt' K(t-t')f[x_P(t')] - \mu_M x_M(t),
\EE
for the drift terms, as well as
\BE\label{eq:kramers2}
B_{P,P}(t,t',\bx) &=& \big(\alpha_P x_M(t) + \mu_P x_P(t) \big) \delta(t-t'),\nonumber \\
B_{P,M}(t,t',\bx) &=& B_{M,P}(t,t',\bx) = 0, \nonumber \\
B_{M,M}(t,t',\bx) &=& \left(\mu_M x_M(t) + \alpha_M\int_{-\infty}^t dt'' K(t-t'')f[x_P(t'')]\right)\delta(t-t'),
\EE
for the correlator of the noise $\boldeta$.
\subsection{Linear-noise approximation and spectra of noise-driven quasi-cycles}
\label{appendix:gene_ps}

For the parameters discussed in the main paper the deterministic dynamics, obtained in the limit $N\to \infty$, is seen to converge to a fixed point $(x_P^*,x_M^*)$. The stochastic system exhibits noise-driven quasi-cycles. In order to predict the spectral properties of these cycles we follow the standard procedure outlined above, see also \cite{mckane}, and write $ x_P= x_P^* + \xi_P/\sqrt{N}$ and $x_M = x_M^* + \xi_M/\sqrt{N}$. Substituting this in Eq. (7) of the main paper, we then perform an expansion in powers of $N^{-1/2}$.  One obtains
\be\label{eq:linear}
\dot\boldxi (t)= \int_{-\infty}^t dt'~ \underline{\underline J}(t-t',\bx^*)\boldxi(t')+\boldeta(t),
\ee
where $\underline{\underline J}(t-t',\bx^*)$ is the (functional) Jacobian of the deterministic dynamics, evaluated at the deterministic fixed point. Specifically we have $J_{\alpha,\beta}(t,t',\bx^*)=\left.\frac{\delta F_\alpha(t,\bx)}{\delta x_\beta(t')}\right|_{\text{FP}}$, i.e
\BE
J_{P,P}(t-t',\bx^*)&=&-\mu_P\delta(t-t'), \nonumber \\
J_{P,M}(t-t',\bx^*)&=&\alpha_P\delta(t-t'), \nonumber \\
J_{M,P}(t-t',\bx^*)&=& \alpha_M K(t-t')f'(x_P^*),\nonumber \\
J_{M,M}(t-t',\bx^*)&=&-\mu_M \delta(t-t').
\EE
with $f'(x_P)$ the derivative of $f(x_P)$ with respect to $x_P$. We note that these matrix elements are time-translation invariant, i.e. they are functions of $t-t'$ only. We also point out that $K(t-t')=0$ for $t'>t$ by causality.

Carrying out a Fourier transform (with respect to $t-t'$) of Eq. (\ref{eq:linear}) we have
\be
\underline{\underline M}(\omega)\widetilde{\boldxi}(\omega) = \widetilde{\boldeta}(\omega),
\ee
where
\be
\underline{\underline M}(\omega) = i\omega \underline{\underline I} - \widetilde{\underline{\underline J}}(\omega,\bx^*), 
\ee
and where $\underline{\underline I}$ is the $2\times 2$ identity matrix. For this model we have that,
\be
\underline{\underline M}(\omega) = \left(\begin{array}{cc} i\omega +\mu_P & -\alpha_P  \\
-\alpha_M\widetilde K(\omega)f'[x_P^*]&i \omega  +\mu_M \end{array} \right) .
\ee
At the deterministic fixed point we have
\BE
B_{P,P}(t-t',\bx^*) &=& \big(\alpha_P x_M^* + \mu_P x_P^* \big) \delta(t-t'), \nonumber \\
B_{P,M}(t-t',\bx^*) &=& B_{M,P}(t,t',\bx^*) = 0, \nonumber \\
B_{M,M}(t-t',\bx^*) &=& \left(\mu_M x_M^* + \alpha_M f[x_P^*)]\right)\delta(t-t').
\EE
We will denote the Fourier transforms of these matrix elements by $\widetilde B_{\alpha,\beta}(\omega)$.

The matrix of power spectra, $\underline{\underline S} = \avg{\boldxi(\omega)\boldxi^\dagger(\omega)}$, is then obtained as 
\be\label{eq:allspec}
\underline{\underline S} (\omega)= \underline{\underline M}^{-1}(\omega)\widetilde{\underline{\underline B}}(\omega) \underline{\underline M}^{\dagger-1}(\omega).
\ee
The diagonal elements of $\underline{\underline S}$ are known as the power spectra, $P_\alpha(\omega)$. We find
\be\label{eq:genespec}
\begin{split}
P_M(\omega) &= \frac{(\mu_P^2 + \omega^2)(\alpha_M f[x_P^*] + \mu_M x_M^*) + (\alpha_Px_M^* + \mu_P x_P^*)(\alpha_M f'[x_P^*] |\widetilde K(\omega)|)^2}{(\mu_M\mu_P - \omega^2 - \alpha_M\alpha_P f'[x_P^*]\text{Re}[\widetilde K(\omega)])^2 + (\omega(\mu_M+\mu_P) - \alpha_M\alpha_P f'[x_P^*] \text{Im}[\widetilde K(\omega)])^2},\\
P_P(\omega) &= \frac{\alpha_P^2(\alpha_M f[x_P^*] +\mu_M x_M^*) + (\mu_M^2+\omega^2)(\alpha_P x_M^*+\mu_P x_P^*))}{(\mu_M\mu_P - \omega^2 - \alpha_M\alpha_P f'[x_P^*]\text{Re}[\widetilde K(\omega)])^2 + (\omega(\mu_M+\mu_P) - \alpha_M\alpha_P f'[x_P^*] \text{Im}[\widetilde K(\omega)])^2} .
\end{split}
\ee
 
For constant delay, $\tau_0$, we have $K(\tau) = \delta(\tau-\tau_0)$, i.e. $\widetilde K(\omega) = e^{-i\omega \tau_0}$, and we recover the result of \cite{galla}.

\section{SIR-model with delayed recovery}
\label{appendix:sir}
\subsection{Gaussian approximation}
In the SIR-model with delayed recovery \cite{andyalan}, defined in the main paper, there are two independent types of particles, $S$ and $I$. The number of recovered individuals follows from the constant overall population size. We write
\be
\bx = \left(\begin{array}{c} S\\I\end{array}\right) .
\ee
In the model there are three reactions, two of which are delay reactions, labelled $i=2,3$. The reaction rates are
\BE
\br[\bx(t)] = \left( \begin{array}{c} \mu(1-S-I)\\ \chi\beta S I \\ (1-\chi)\beta S I\end{array} \right) .
\EE
The immediate effects of the reactions (i.e. at the time they are triggered) are given by
\BE
\underline{\underline v} = \left (\begin{array}{cc} 1&0\\-1&1\\-1&1\end{array}\right),
\EE
 and the delayed effects on particle numbers are described by
\BE
\underline{\underline w} = \left (\begin{array}{cc}0&0\\0&-1\\1&-1\end{array}\right) .
\EE
The distributions of delay times are $K_2=K(\tau)$ and $K_3= Q(\tau)$, with $K(\tau)$ and $Q(\tau)$ as defined in the main paper. It turns out to be convenient to define $\olK (\tau) = \chi K(\tau)$ and $\olQ (\tau) = (1-\chi) Q(\tau)$.

Putting all this together, and applying Eqs. (3, 4, 5) of the main paper we have
\BE
F_{S}(t,\bx)&=&-\beta S(t)I(t) + \mu(1-S(t)-I(t))+\beta\int d\tau ~\olQ(\tau)S(t-\tau)I(t-\tau), \nonumber \\
F_I(t,\bx) &=& \beta S(t)I(t) -\beta\int d\tau~ [\olQ(\tau)+\olK(\tau)]S(t-\tau)I(t-\tau),\label{eq:sirdrift}
\EE
as well as
\BE
B_{S,S}(t,t',\bx) &=& \bigg\{\left(\beta S(t)I(t) +\mu(1-S(t)-I(t)) +\beta\int d\tau ~\olQ(\tau)S(t-\tau)I(t-\tau)\right) \delta(t-t') \nonumber \\
&&- \beta \olQ(t-t')S(t')I(t') -\beta\olQ(t'-t)S(t)I(t)\bigg\},\nonumber \\
B_{S,I}(t,t',\bx) &=& \bigg\{\left(-\beta S(t)I(t) - \beta\int d\tau~ \olQ(\tau)S(t-\tau)I(t-\tau)\right)\delta(t-t')\nonumber \\
&&+  \beta \olQ(t-t')S(t')I(t') + \beta \Big(\olQ(t'-t)+\olK(t'-t)\Big)S(t)I(t) \bigg\},\nonumber \\
B_{I,S}(t,t',\bx) &=& \bigg\{\left(-\beta S(t)I(t) - \beta\int d\tau~ \olQ(\tau)S(t-\tau)I(t-\tau)\right)\delta(t-t')\nonumber \\
&&+  \beta \olQ(t'-t)S(t)I(t) + \beta \Big(\olQ(t-t')+\olK(t-t')\Big)S(t')I(t') \bigg\},\nonumber \\
B_{I,I}(t,t',\bx) &=& \bigg\{\left(\beta S(t)I(t)+\beta\int d\tau ~\Big(\olQ(\tau)+\olK(\tau)\Big)S(t-\tau)I(t-\tau)\right) \delta(t-t') \nonumber \\
&&- \beta \Big(\olQ(t-t')+\olK(t-t')\Big)S(t')I(t') - \beta \Big(\olQ(t'-t)+\olK(t'-t)\Big)S(t)I(t) \bigg\}.\label{eq:sircorr}
\EE
\subsection{Linear-noise approximation and spectra of quasi-cycles}
\label{appendix:sir_ps}
Concentrating again on parameter ranges in which the deterministic model, obtained for $N\to\infty$, has a fixed point $(S^*, I^*)$, we decompose $S = S^* + \xi_S/\sqrt{N}$ and $I = I^* + \xi_I/\sqrt{N}$. Within a systematic expansion in powers of $N^{-1/2}$ this allows us to substitute $S(t)\to S^*$ and $I(t)\to I^*$ in Eqs. (\ref{eq:sirdrift}) and (\ref{eq:sircorr}).

As in the model of gene regulation we next carry out a Fourier transform, and find
\be
\underline{\underline M}(\omega)\widetilde{\boldxi}(\omega) = \widetilde{\boldeta}(\omega)
\ee
where
\be
\underline{\underline M}(\omega) = i\omega \underline{\underline\id} - \widetilde{\underline{\underline J}}(\omega,\bx^*). 
\ee
For the SIR-model we have 
\be
\underline{\underline M}(\omega) = \left(\begin{array}{cc} i\omega + \mu + \beta[1- \widetilde Q(\omega)]I^*  & \mu + \beta[1- \widetilde Q(\omega)]S^*  \\
-\beta[1-\widetilde K(\omega)  -\widetilde Q(\omega) ]I^* &i \omega  -\beta[1-\widetilde K(\omega)-\widetilde Q(\omega)]S^* \end{array} \right).
\label{eq:M-matrix}
\ee
The power spectrum for the infectives, $I$, is given by $P_I(\omega) = \avg{|\widetilde \xi_I(\omega)|^2}$, and is found as
\BE
P_I(\omega) =& \frac{1}{|\text{det}~\mathbb{M}(\omega)|^2}\big(|M_{SS}(\omega)|^2\widetilde B_{II}(\omega) - M_{SS}(\omega)M_{IS}^*(\omega)\widetilde B_{IS}(\omega) \nonumber \\
&-M_{IS}(\omega)M_{SS}^*(\omega)\widetilde B_{SI}(\omega) + |M_{SI}(\omega)|^2\widetilde B_{SS}(\omega) \big).\label{eq:sirpowerspectrum}
\EE
For a specific instance of the model, i.e. for a specific choice of the delay kernel $H(\tau)$  it is then a matter of first finding $K(\tau), Q(\tau)$ and $\chi$, see main paper. A Fourier transform then gives $\widetilde K(\omega)$, $\widetilde Q(\omega)$. The deterministic fixed point, $S^*$ and $I^*$, can be found analytically for many kernels, as can the Fourier transforms, $\widetilde K(\omega)$ and $\widetilde Q(\omega)$ . Eq. \eqref{eq:sirpowerspectrum} can then be evaluated to obtain an analytical prediction for the power spectrum of quasi-cycles of the number of infectives in the population within the linear-noise approximation.
 
\begin{figure}[t!!!!]
\vspace{0.5em}
\centerline{\includegraphics[angle=270, width=0.45\textwidth]{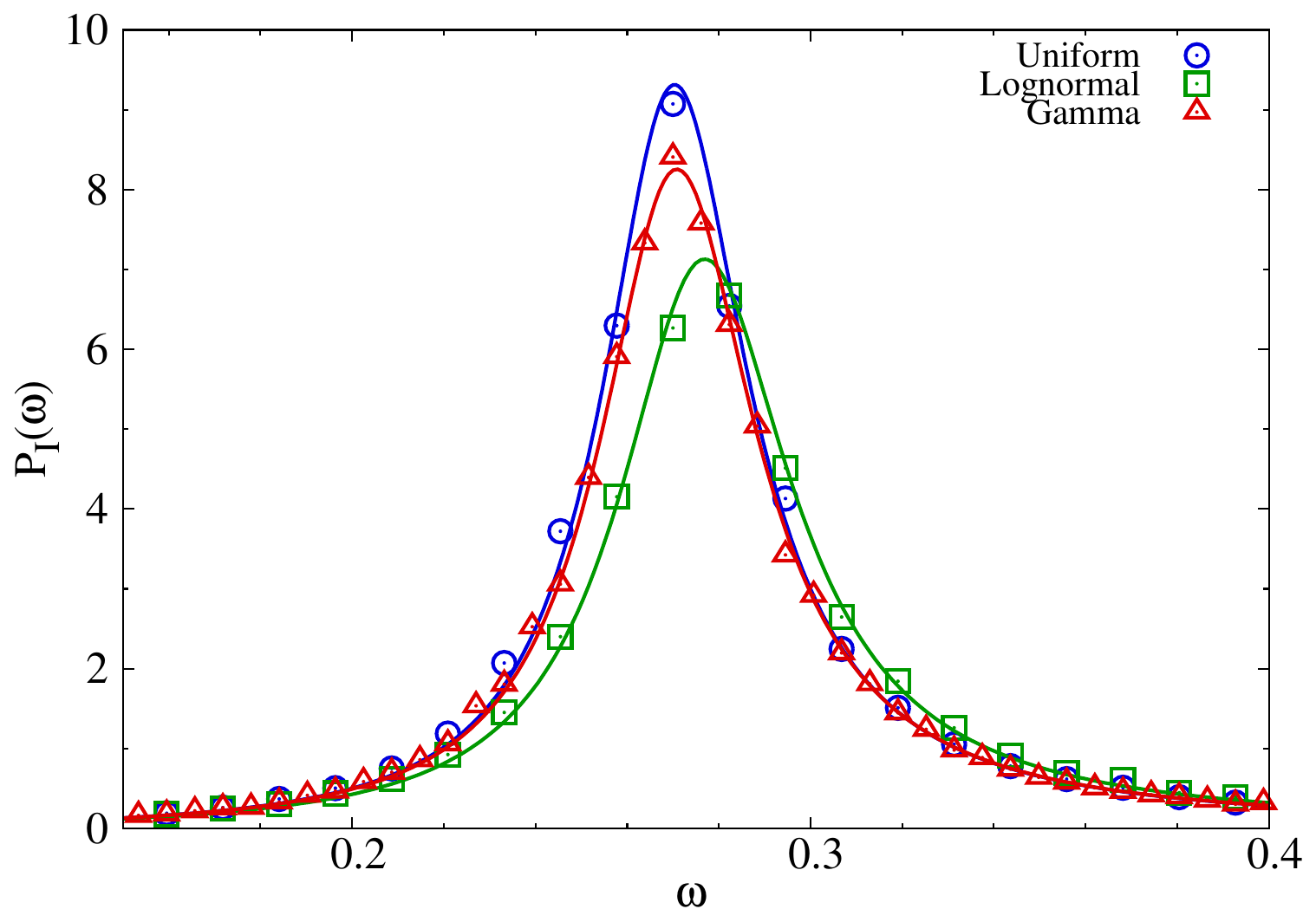}}
\vspace{1em} 
\caption{(Colour on-line) Power spectra of quasi-cycles of the number of infectives in the SIR model with recovery times drawn from a uniform distribution, a lognormal distribution and a $\Gamma$-distribution respectively. All distributions are chosen to have the same mean, $\avg{\tau}=1$, and variance $\avg{\tau^2}-\avg{\tau}^2=1/L$. Parameters are the same as in Fig. 3 of the main paper with $L = 4$.  Markers are from simulations (averaged over $800-1000$ samples), lines are from the theory in the LNA. }
\label{fig:lognormalflat}
\end{figure}

In the main paper we show results for the SIR model with $\Gamma$-distributed delays. The result of Eq. (\ref{eq:sirpowerspectrum}) is general however, and can be evaluated numerically for any choice of the delay kernel. Arguments for various different distributions of for example infectious periods and recovery times have been presented in the literature \cite{kernels}.
In order to confirm the validity of our approach for delay kernels other than the $\Gamma$-distribution we have studied the case of a lognormal delay distribution (mentioned e.g. in Sartwell \cite{kernels}), and a flat delay kernel. Corresponding results are shown in Fig. \ref{fig:lognormalflat}.
 
 \section{Simulations of the chemical Langevin equation for processes with delay}
\label{appendix:simulate}
\subsection{Implementation and basic test}

 In the proceeding sections we have focused on the use of the approximation Eqs. (3, 4, 5) of the main paper in calculating the power spectra. These equations are the analogue of what is known as the chemical Langevin equation for systems without delay. They can also serve as the basis for efficient simulations of the stochastic dynamics. Specifically Eqs. (3, 4, 5) of the main paper provide an approximation of the underlying discrete particle system in terms of a stochastic process with multiplicative Gaussian noise. The chemical Langevin equation can be discretised (in time) relatively straightforwardly, even for delay systems.

In this section we explore numerical approaches based on the chemical Langevin equation, with focus on the model of gene regulation, as defined in the main paper. The corresponding chemical Langevin equation reads (see Eqs. (7, 8) of the main paper):
\BE
\dot x_M(t)&=& \alpha_M \int_{0}^\infty d\tau ~  K(\tau)f[x_P(t-\tau)] - \mu_M x_M(t)+N^{-1/2}\eta_M(t), \nonumber \\
\dot x_P(t) &=& \alpha_P x_M(t) - \mu_P x_P(t)+N^{-1/2}\eta_P(t),  
\EE
where $f[x_P(t)]=[1+(x_P(t)/P_0)^h]^{-1}$, and
\BE
\avg{\eta_M(t) \eta_M(t')} &=& \bigg[\alpha_M \int_{0}^\infty d\tau ~  K(\tau)f[x_P(t-\tau)] + \mu_M x_M(t)\bigg] \delta(t-t'), \nonumber \\
\avg{\eta_P(t) \eta_P(t')} &=& \left[\alpha_P x_M(t) + \mu_P x_P(t)\right] \delta(t-t'), \nonumber \\
\avg{\eta_M(t) \eta_P(t')} &=& 0.
\EE

We have discretised these integro-differential equations using a simple first-order scheme. Care needs to be taken, in principle, when it comes to discretizing the multiplicative noise. Given that Gaussian random numbers are used, any scheme with a finite time step will result in a finite probability that the correlation matrix of $\eta_M$ and $\eta_P$ develops negative eigenvalues, which renders the generation of noise invalid. Specific schemes are available to deal with such situations in non-delay systems with one degree of freedom, see in particular \cite{moro, dornic}. In our implementation we have used a naive first-order scheme, and we have not encountered any problems, presumably because the dynamics operates sufficiently far from extinction of the individual types of molecules, i.e. their particle numbers do not assume values near zero.

\begin{figure}[t!!!!]
\vspace{0.5em}
\centerline{
\includegraphics[angle=270, width=0.32\textwidth]{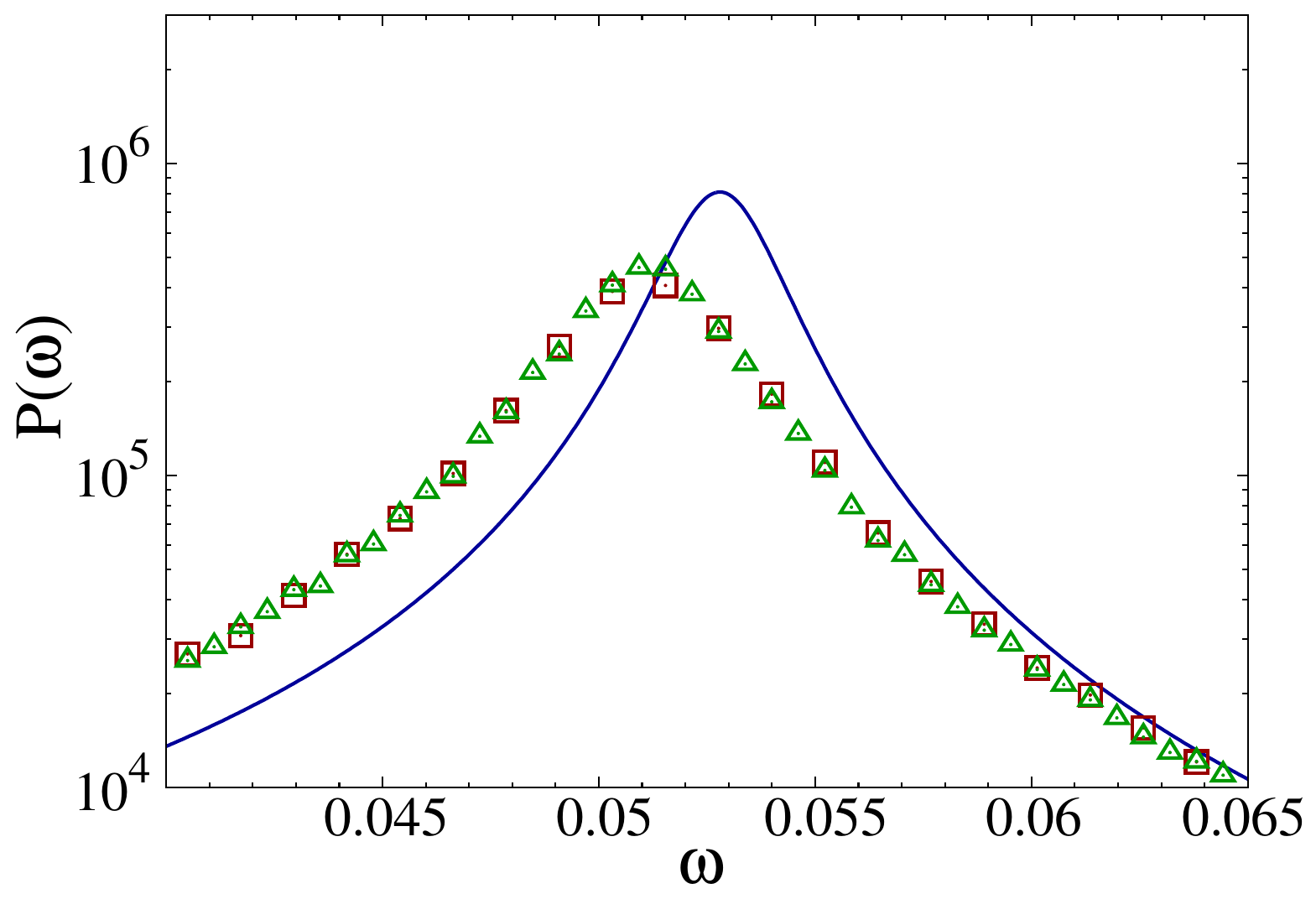}
\includegraphics[angle=270, width=0.32\textwidth]{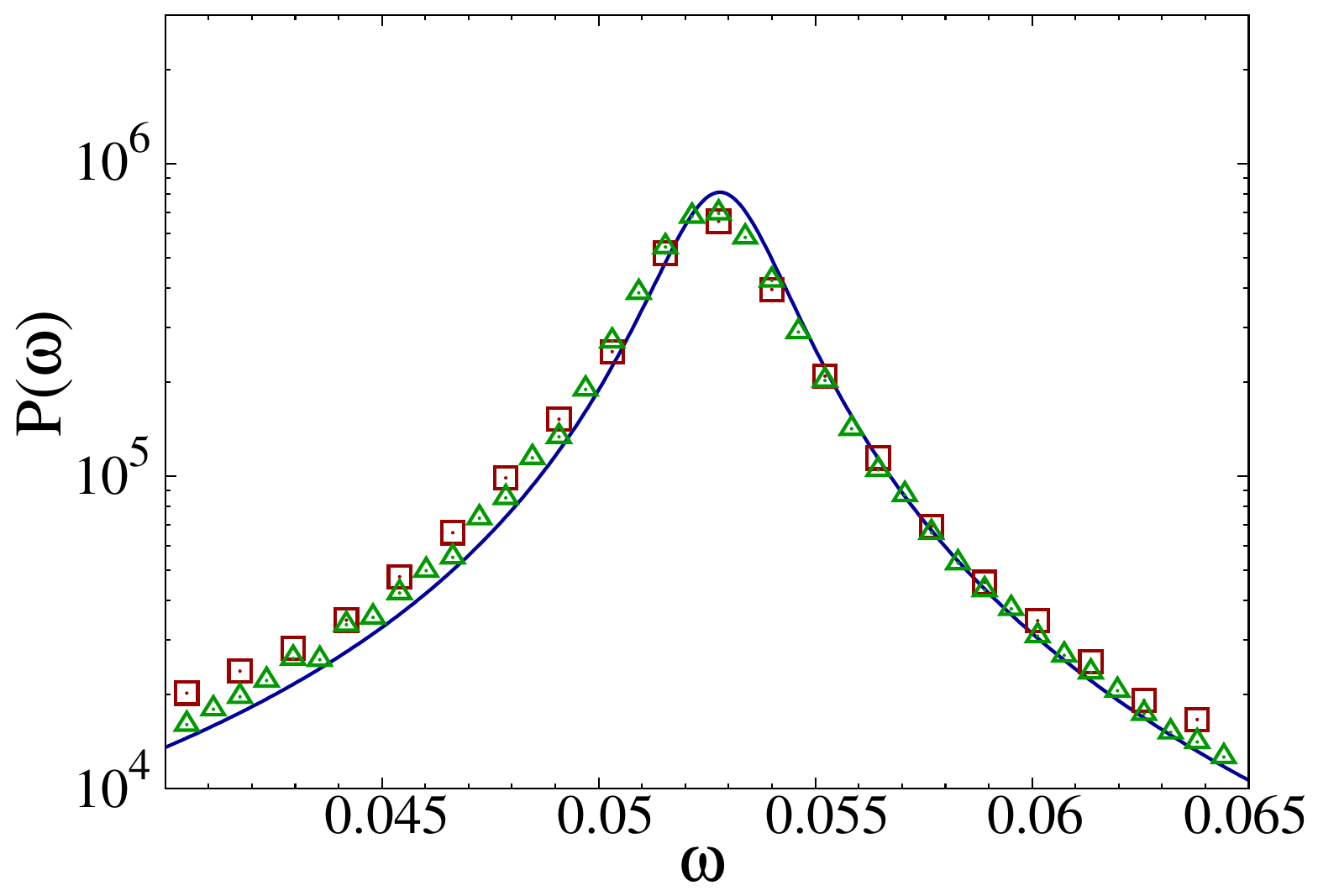}
\includegraphics[angle=270, width=0.32\textwidth]{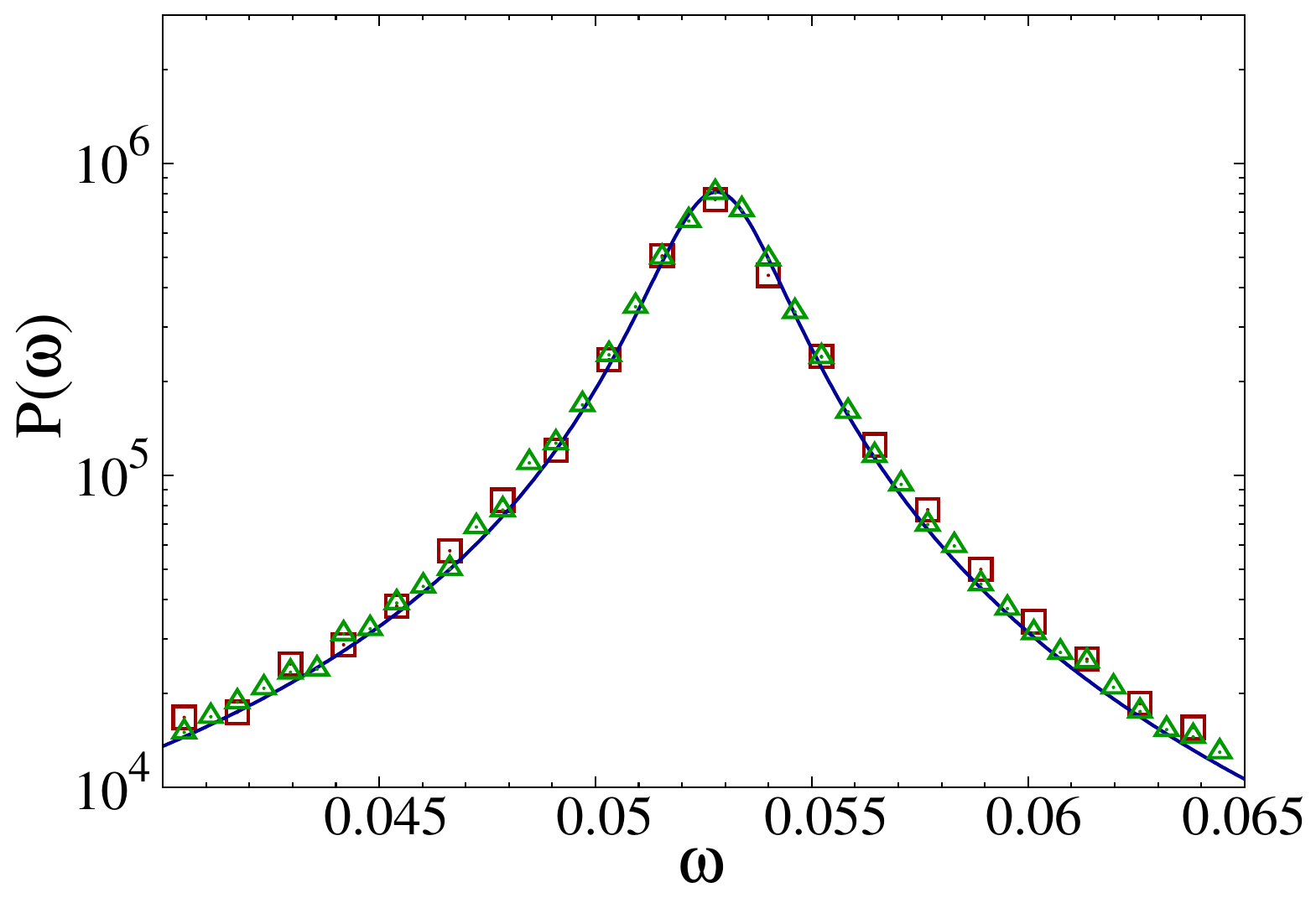}
}
\vspace{1em} 
\caption{(Colour on-line) Power spectra of quasi cycles of protein numbers in the gene regulation model. Squares show results from Gillespie simulations of the original process, triangles from a discretization of the chemical Langevin equation. Solid lines are from the analytical LNA-calculation. Simulations are for $\kappa=16$. Scale of the system size is $N=100$ (left-hand panel), $N=1000$ (central panel) and $N=5000$ (right-hand panel). All other model parameters are as in Fig. 1 of the main paper. Simulation data is averaged over $100-1000$ independent runs, depending on system size.}
\label{fig:lspec}
\end{figure}

As a test we have carried out simulations of the delay Langevin dynamics to reproduce the power spectra of quasi-cycles as seen in simulations of the original discrete particle model. Results are shown in Fig. \ref{fig:lspec}. Unsurprisingly simulations of the Langevin dynamics agree accurately with Gillespie simulations of the discrete process and with theoretical results from the LNA when particle numbers are sufficiently large. At lower system sizes the LNA may be less accurate (see left-hand panel of Fig. \ref{fig:lspec}). Simulations of the chemical Langevin equation may however still agree with data from the original dynamics in such cases.

\subsection{Relative running times}
Typical running times of simulations are shown in Table \ref{tab:runtimes}, all times indicated are real times. As confirmed in the table the running time of simulations of the chemical Langevin equation does not depend on the system size. This is not surprising, the system size enters only through the magnitude of the noise term in the multiplicative process. It does not affect the first-order numerical algorithm used to solve the stochastic integro-differential equation, so the computing time is independent of the system size. The running time of the Gillespie algorithm does however depend on  the typical system size. In the standard Gillespie algorithm, applied to systems without delay interactions, the typical time increment after each Gillespie step scales as $N^{-1}$, so the total running time until a fixed final time increases linearly in $N$. The situation for systems with distributed 
delays is more complicated, as a list of initiated delay reactions needs to be kept and updated during the process of the simulation. More crucially reactions on this list need to be ordered according to the times at which they are to be executed. The typical number of entries in the list grows linearly in $N$, and so the algorithm requires additional computing time.

We have not carried out a systematic analysis of the performance of our simulations of chemical Langevin equations, and we have not optimised the procedures used for the modified Gillespie algorithm and for the Langevin simulations. The data shown in Table \ref{tab:runtimes} however confirms that the amount of time required to carry out Gillespie simulations of the delay dynamics grows with the system size (presumably supra-linearly), and that the run time required for simulations of the Langevin dynamics is independent of $N$. In our implementation the modified Gillespie algorithm is hence preferable for very small system sizes. As the number of particles is increased simulating the chemical Langevin equation quickly outperforms the Gillespie approach.
\begin{table}[t!!!!]
\begin{center}
\begin{tabular}{|c|c|c|}
\hline
~~~$N$~~~& ~~~Gillespie~~~ & ~~~Langevin~~~   \\
\hline
100 & 3.5s & 24.9s  \\
\hline
1000 & 41.9s & 24.8s \\
\hline
~~~5000 ~~~& 5438.7s & 24.8s \\
\hline
 \end{tabular}
\caption{\label{tab:runtimes} Estimates of running times (real time) to complete $20$ runs of the gene regulation model up to time $t=10^6$ for different scales $N$ of the system size ($\kappa=16$, all other model parameters as in Fig. 1 of the main paper). We here use a modified Gillespie method for the original discrete particle process with delay, and a first-order integration scheme for the chemical Langevin equation. We stress that neither algorithm has been fully optimised in our implementation.}
\end{center}
\vspace{-0.6cm}
\end{table}

\subsection{Stationary distributions}
As a further application of the chemical Langevin equation and the LNA we have studied the stationary probability distribution, $P(n_M, n_P)$ of the number of protein and mRNA molecules in the model of gene regulation. In the LNA we have $n_M=Nx_M^*+\sqrt{N}\xi_M$, and similarly for the protein. The distribution of $n_M$ and $n_P$ is Gaussian, with a mean given by the deterministic fixed point, $(Nx_M^*, Nx_P^*)$.

Keeping in mind that the auto-correlation functions of $\xi_M(t)$ and $\xi_P(t)$, are the Fourier transforms of the spectra $P_M(\omega)$ and $P_P(\omega)$ respectively, we have $\avg{\xi_M(t)\xi_M(t)}=(2\pi)^{-1}\int_{-\infty}^{\infty} d\omega P_M(\omega)$ in the stationary state (i.e. at asymptotic times $t$), and similarly for $\xi_P$. Expressions for $P_M(\omega)$ and $P_P(\omega)$ are given in Eq. (\ref{eq:genespec}). The asymptotic equal-time covariance, $\avg{\xi_M(t)\xi_P(t)}$, can be calculated from the cross-spectrum $\avg{\widetilde \eta_M(\omega)\widetilde \eta_P^*(\omega)}$, the latter in turn is obtained from Eq. (\ref{eq:allspec}). The remaining integrals over $\omega$ are performed numerically.

\begin{figure}[t!!!!]
\vspace{0.5em}
\centerline{\includegraphics[angle=270, width=0.4\textwidth]{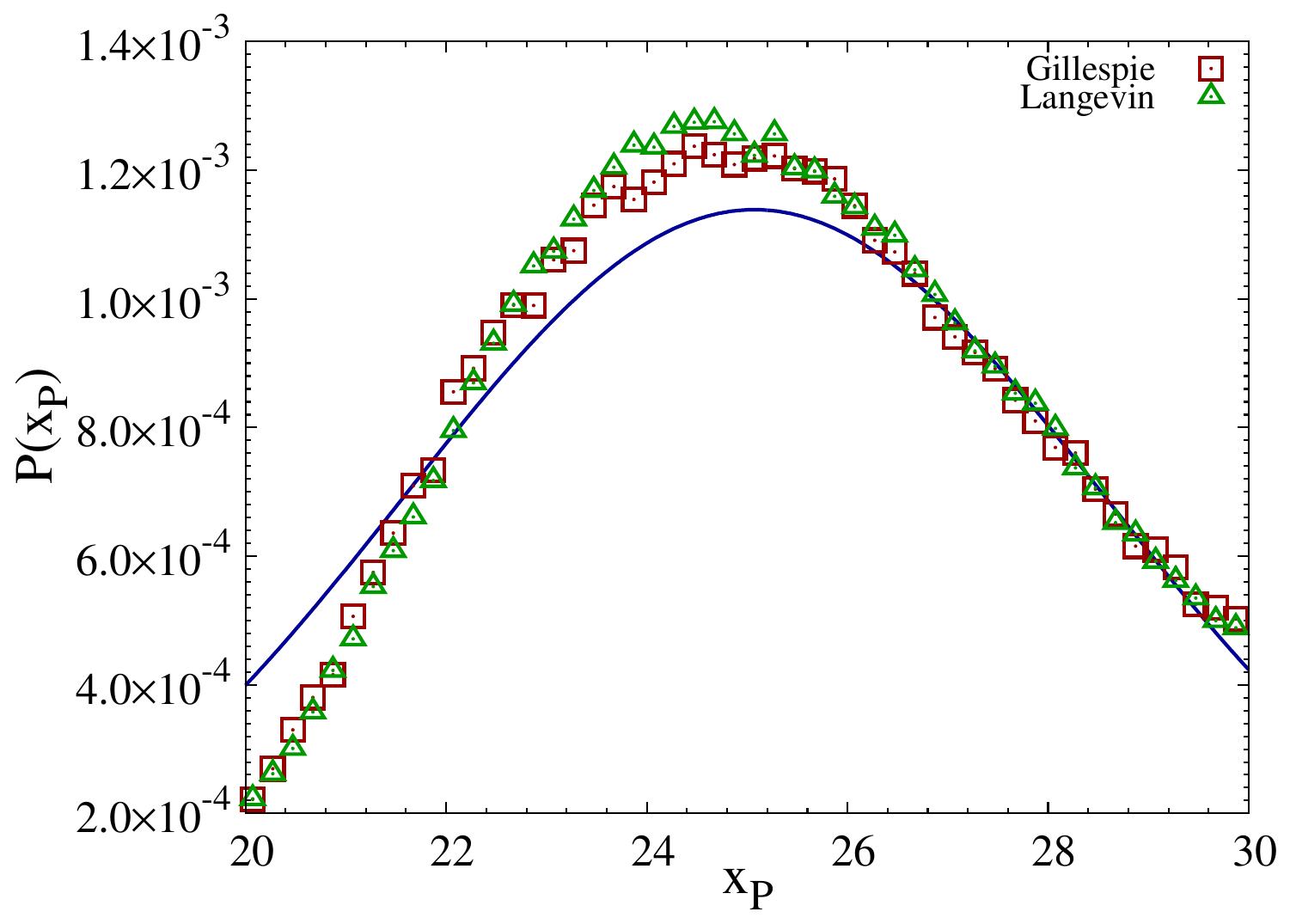}
~~~~~~~~~~~ \includegraphics[angle=270, width=0.4\textwidth]{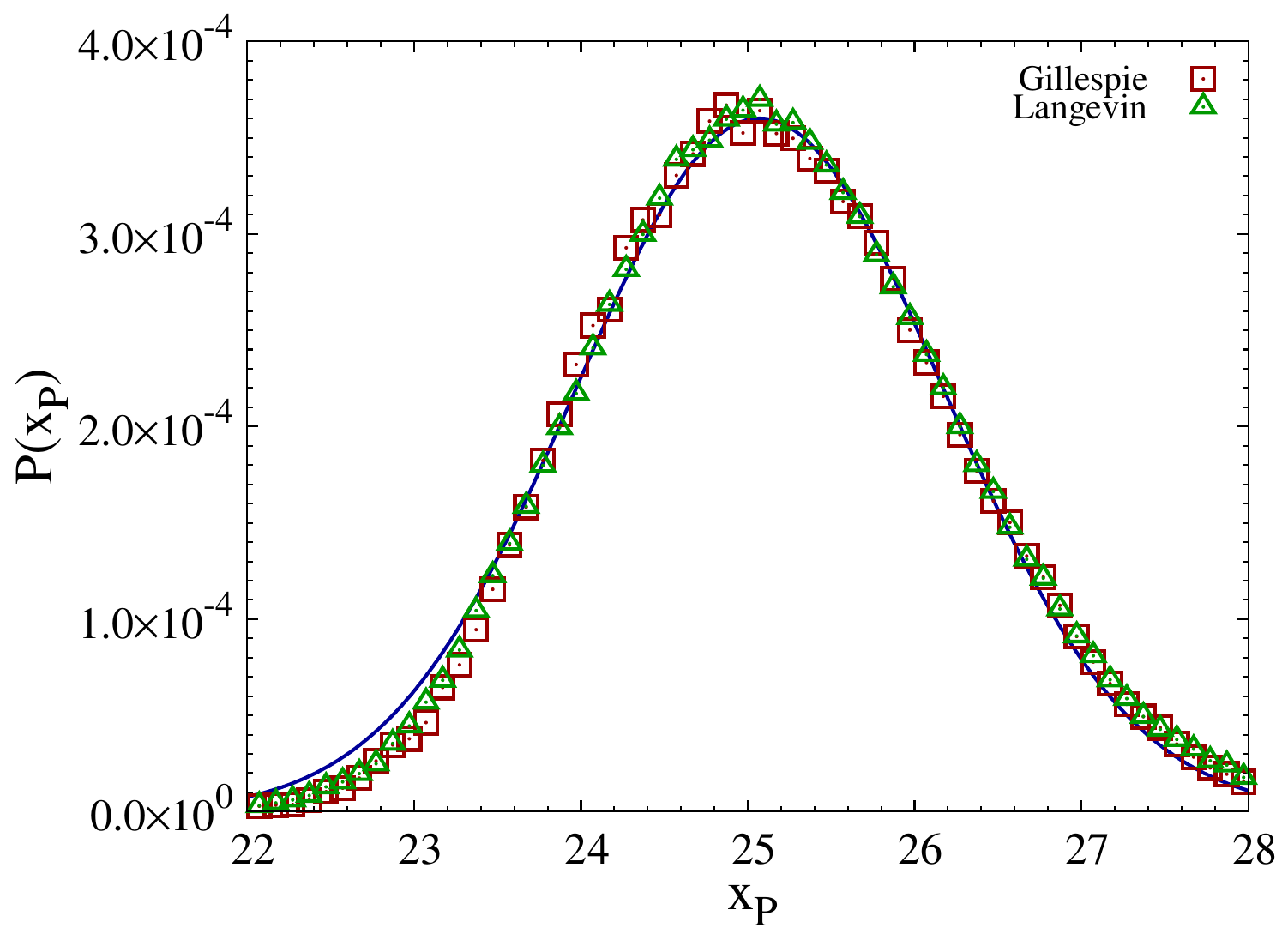}
\vspace{0.5em}}
\caption{(Colour on-line)  Stationary distribution of the number of protein molecules in the model of gene regulation ($\kappa=16$, all other model parameters as in Fig. 1 of the main paper). The panel on the left shows results for $N=100 $, the right-hand panel is for $N=1000$. Squares show results from simulations of the original microscopic process (100 samples), triangles are from a numerical integration of the chemical Langevin equation with delay, Eq. (\ref{eq:kramers1}, \ref{eq:kramers2}), data from 100 samples. Solid lines show results obtained in the linear noise approximation. }
\label{fig:geneprob}
\end{figure}

We compare these semi-analytical predictions with data from Gillespie simulations and with results from a numerical integration of the chemical Langevin equation in Fig. \ref{fig:geneprob}. The LNA works well for large systems (right-hand panel). At small particle numbers the chemical Langevin equation is a better approximation of the original process than the LNA.

\begin{figure}[h!]
\vspace{0.5em}
\centerline{\includegraphics[angle=270, width=\textwidth]{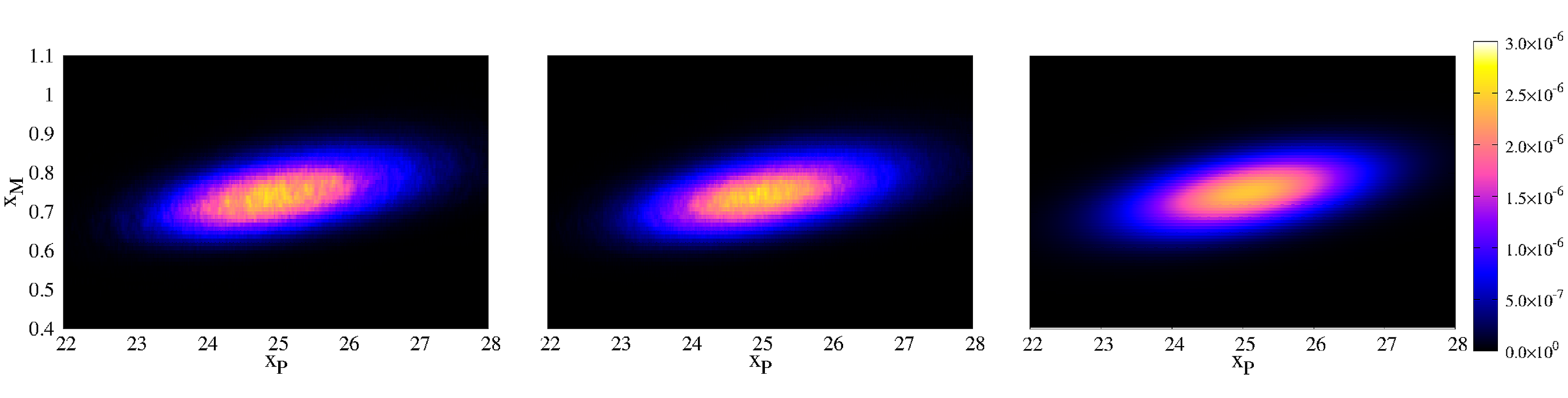}
 }

\caption{(Colour on-line)  Stationary distribution of the number of protein molecules and of the number of in the mRNA molecules in the model of gene regulation ($\kappa=16$, $N=1000$, all other model parameters as in Fig. 1 of the main paper). The left-hand panel shows data from Gillespie simulations of the original process, the central panel is from a numerical integration of the chemical Langevin dynamics, and the right-hand panel shows semi-analytical results from the LNA. Simulations are averaged over $100$ runs.}
\label{fig:geneprob2}
\end{figure}

In Fig. \ref{fig:geneprob2}, finally, we show results for the joint stationary distribution, $P(n_M, n_P)$, as obtained from Gillespie simulations (left-hand panel), from the chemical Langevin equation (centre) and from the LNA (right-hand panel).  

 \subsection{Further comments: analytical tractability of chemical Langevin equations with delay}
 The dynamics in the LNA is by definition linear, and hence it can be solved exactly, at least in principle. For stochastic delay systems there will generally be non-Markovian terms coupling back in time, and the noise is coloured, but in the linear approximation closed expressions can be derived for the statistics of fluctuations in the stationary state, as discussed above. The chemical Langevin equation for delay systems {\em before} the LNA is made is generally non-linear, and as such full analytical solutions will not normally be possible. We note that this is the case as well for non-delay systems, and mainly a consequence of the non-linearities and the multiplicative nature of the resulting noise. Delay and correlation structures in the noise complicate these matters even further. 
 
 A further point concerns the analytical calculation of quantities beyond the statistics of the stationary distribution or correlation functions. Objects of interest might for example be first-passage or extinction times \cite{gardiner,redner}. In systems without delay these are usually computed starting from a backward Fokker-Planck or master equation. Our formalism does not use Fokker-Planck or master equations, previous work on stochastic delay systems is frequently based on equations which do not fully close \cite{frank, bratsun}. So it is hard to judge how much can be said analytically. We do however expect that, at least for sufficiently large systems, simulations based on the chemical Langevin equation are the most efficient route towards measuring such quantities.
 
 A further remark concerns the derivation of an analogue of the so-called Liapunov equation for correlation functions in the LNA. These are closed form equations governing the time-dependence of the {\em equal-time} correlation matrix $C_{\alpha, \beta}(t,t)$ of fluctuations (where $\alpha$ and $\beta$ label different species of particles). These equations can be formulated for the linearised Gaussian dynamics of systems without delay (see e.g. \cite{risken}), and involve only one-time objects, such as $C_{\gamma,\delta}(t,t)$. We have derived similar equations for delay systems (from the LNA), but find that they do not close in terms of one-time objects. Instead the time derivative, $dC_{\alpha,\beta}(t,t)/dt$ involves terms of the type $C_{\gamma,\delta}(t,t')$, with $t'\leq t$, and the formalism becomes more involved.



\begin{thebibliography}{99}
\bibitem{remark-1}For $\Delta>0$ the model dynamics can lead to unphysical negative particle numbers with finite probability. In the limit $\Delta \to 0$, taken at the end of the calculation, this probability vanishes.
\bibitem{remark0} We assume that any reaction induces changes in particle numbers at (at most) one later time. Extensions to cases in which changes occur at multiple later times are relatively straightforward.
\bibitem{remark1} In the Markovian case ($w_{i,\alpha}^\tau=0$) our result reduces to that obtained from a Kramers-Moyal expansion or from Kurtz' theorem. The Gaussian noise $\boldeta(t)$ is then white, and the retarded interaction term in $F_\alpha$ is absent.
\bibitem{remark2} The scaling of time (and hence frequencies $\omega$) in \cite{andyalan} is different from the conventions we use here.
\end{thebibliography}
\end{document}